\documentclass[preprint,aps,showpacs,nofootinbib,preprintnumbers,amsmath,amssymb]{revtex4-1}
\usepackage{amsfonts}
\usepackage{amssymb}
\usepackage{}
\usepackage{epsfig}
\usepackage{graphicx}
\usepackage{subfigure}
\usepackage{dcolumn}
\usepackage{bm}
\usepackage[usenames ,dvipsnames]{xcolor}
\usepackage{slashed}

\begin{document}

\title{Loop-induced Neutrino Masses: A Case Study}

\author{Chao-Qiang~Geng$^{1,2,3}$\footnote{geng@phys.nthu.edu.tw},
Da~Huang$^{2}$\footnote{dahuang@phys.nthu.edu.tw}
and
Lu-Hsing Tsai$^{2}$\footnote{lhtsai@phys.nthu.edu.tw}}
  \affiliation{$^{1}$Chongqing University of Posts \& Telecommunications, Chongqing, 400065, China\\
  $^{2}$Department of Physics, National Tsing Hua University, Hsinchu, Taiwan\\
  $^{3}$Physics Division, National Center for Theoretical Sciences, Hsinchu, Taiwan
}

\date{\today}

\begin{abstract}
We study the cocktail model in which the Majorana neutrino masses are generated by the so-called ``cocktail'' three-loop diagrams with the dark matter particle running in the loops. In particular, we give the correct analytic expressions of the neutrino masses in the model by the detailed calculation of the cocktail diagrams. Based on the reliable numerical calculation of the loop integrals, we explore the parameter space which can give the correct orders of neutrino masses while satisfying other experimental constraints, such as those from the neutrinoless double beta decay, low-energy lepton flavor violation processes, electroweak precision tests, and collider searches. As a result, the large couplings and the large mass difference between the two singly-charged (neutral) scalars are required.
\end{abstract}

\maketitle
\section{Introduction}\label{intro}
The small but non-zero masses and mixings of the neutrinos have been found via the neutrino oscillation experiments~\cite{NuOscillation}, while 
dark matter (DM) has been established by astrophysical observations~\cite{RotationCurve,Planck,BulletCluster}. Both phenomena cannot be explained within the Standard Model (SM) of particle physics, directly pointing to the existence of new physics. 

One possible explanation of the tiny neutrino masses is the canonical seesaw mechanisms, where the masses are generated at tree level by introducing right-handed neutrinos~\cite{TypeI-Seesaw} or a Higgs triplet~\cite{TypeII-Seesaw} or fermion triplets~\cite{TypeIII-Seesaw}. Unfortunately, such new particles are predicted too heavy to be studied at the current colliders. Another idea is to promote the neutrino mass generation to loop levels~\cite{RadiativeSeesaw}, where the smallness of the neutrino masses is attributed to the loop suppression and the masses of the new particles are naturally of ${\cal O}$(100 $\sim$ 1000)~GeV or even smaller so that the phenomenology can be very rich. In particular, the discrete symmetries imposed on some models play an extra role to guarantee the stability of DM~\cite{Neutrino_DM}, resulting  in a common origin of neutrino masses and DM. The cocktail model~\cite{Gustafsson:2012vj} is one recent example along this line of thinking, in which the Majorana neutrino mass terms first appear at three-loop level via the so-called ``cocktail'' diagrams, while DM is identified as a neutral $Z_2$-odd particle running in the loops. It is interesting to note that the model naturally predicts the normal hierarchy form of the neutrino mass matrix.

However, the detailed derivations of the formula for the neutrino masses from the cocktail diagrams  were not given in Ref.~\cite{Gustafsson:2012vj}. 
In this paper, we present the full form of the neutrino mass formula in this model. Moreover, with the explicit analytic calculation of the relevant Feynman diagrams and the reliable numerical integration, we explore the parameter space which can give the required values of the neutrino masses while satisfying all of the other constraints from the neutrinoless double beta ($0\nu\beta\beta$) decay, low-energy lepton flavor violation (LFV) processes, electroweak precision tests (EWPTs), and collider searches.

The paper is organized as follows. In Sec.~\ref{CocktailModel}, we show the particle content in the cocktail model and the relevant part of the Lagrangian. In Sec.~\ref{NuPhys}, we examine the neutrino mass matrix by considering the current neutrino oscillation data and the $0\nu\beta\beta$ constraint. We then discuss the constraints from LFV processes, EWPTs, and DM  and collider
searches in Sec.~\ref{flavor},~\ref{EWPT}, and \ref{DM-collider}, respectively. Our numerical exploration of the parameter space is carried on in Sec.~\ref{NumericalRes}. A short summary is given in Sec.~\ref{Conclusion} . In Appendix~\ref{Details}, the analytical calculation details of the cocktail diagrams are presented.

\section{The Cocktail Model for Neutrino Masses}\label{CocktailModel}
Besides the SM fields and symmetries, two $SU(2)_L$ singlet scalars, $S^+$ and $\rho^{++}$, and a scalar doublet $\Phi_2$ are introduced,
and  an exact $Z_2$ symmetry is imposed. Under $Z_2$, $S^+$ and $\Phi_2$ are odd, while $\rho^{++}$ and all the SM fields are even. 
After the electroweak (EW) symmetry breaking, the $Z_2$ symmetry keeps so that the lightest $Z_2$-odd state remains stable and becomes a DM particle candidate. The particle content of the new physics sector is summarized in Table~\ref{ParticleContent},
\begin{table}\caption{New Physics Sector Particle Content of the Cocktail Model}
\begin{tabular}{c|ccc}\label{ParticleContent}
  & $SU(2)_L$ & $U(1)_Y$ & $Z_2$ \\
\hline
$\Phi_2$ & ${\bf 2}$ & 1 & $-$ \\

$S^+$ & ${\bf 1}$ & $2$ & $-$ \\

$\rho^{++}$ & {\bf 1} & $4$ & $+$
\end{tabular}
\end{table}
and the relevant Lagrangian is given by,
\begin{eqnarray}\label{Lagrangian}
-{\mathcal L}_{\rm dark} &=& \frac{\lambda_5}{2}(\Phi_1^\dagger \Phi_2)^2 + \kappa_1 \Phi_2^T i\sigma_2 \Phi_1 S^- + \kappa_2 \rho^{++} S^- S^- \nonumber\\
&& + \xi \Phi^T_2 i\sigma_2 \Phi_1 S^+ \rho^{--} + C_{ab} \overline{\ell^c}_{aR} \ell_{bR} \rho^{++} +{\rm h.c.},
\end{eqnarray}
where, $a$ and $b$ denote the three families of the right-handed leptons $\ell_R$, and $C_{ab}$ are the elements of the Yukawa coupling matrix which is symmetric and complex in general.

After the spontaneous breaking of the EW symmetry, the SM Higgs doublet $\Phi_1$ and inert scalar doublet $\Phi_2$ can be written in the unitary gauge as
\begin{eqnarray}
\Phi_1 = \left( \begin{array}{c}
0 \\
v + \frac{h}{\sqrt{2}}
\end{array} \right), \quad\quad
\Phi_2 = \left( \begin{array}{c}
\Lambda^+ \\ \frac{1}{\sqrt{2}} (H_0+i A_0)
\end{array} \right),
\end{eqnarray}
where $v\approx 174$~GeV is the vacuum expectation value (VEV) of the SM Higgs $\Phi_1$. With a nonzero $\kappa_1$, the charged scalars $\Lambda^+$ and $S^+$ will mix together with an angle $\beta$, leading to two charged mass eigenstates
\begin{eqnarray}
H_1^+ = s_\beta S^+ + c_\beta \Lambda^+, \quad \quad H_2^+ = c_\beta S^+ - s_\beta \Lambda^+ ,
\end{eqnarray}
with $s_\beta ( c_\beta) = \sin\beta(\cos\beta)$. In the mass eigenstate basis, the most useful set of independent variables is the five new scalar masses $m_{\rho,~H_0,~A_0,~H_{1,2}^+}$, the mixing angle $\beta$, and the couplings $\xi$ and $\kappa_2$. All the original parameters defined in the scalar potential in Eq.~(\ref{Lagrangian}) can be solved with these physical parameters.

\begin{figure}[ht]
\centering
\includegraphics[width=0.7\textwidth]{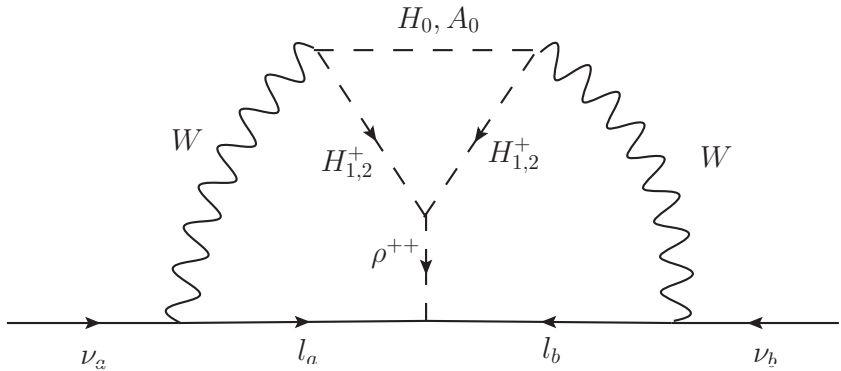}
\caption{Cocktail Diagrams for Neutrino Masses}
\label{FigCocktail}
\end{figure}

The lepton number is explicitly broken in the Lagrangian of Eq.~(\ref{Lagrangian}) by two units, which is the necessary condition to generate the Majorana masses for the three light active neutrinos. However, as pointed in Ref.~\cite{Gustafsson:2012vj}, the leading contribution to the neutrino masses appears at three-loop level via the so-called ``cocktail diagrams'' shown in Fig.~\ref{FigCocktail}. 

In the basis where the charged leptons are in mass eigenstates and the charged current interactions are flavor diagonal, the Majorana neutrino mass matrix elements are given by
\begin{eqnarray}\label{Mass}
(m_\nu)_{ab} &=& (x_a C_{ab} x_b)\frac{s_{2\beta}}{(16\pi^2)^3}({\mathcal A}_1 {\mathcal I}_1+{\mathcal A}_2{\mathcal I}_2),
\end{eqnarray}
with
\begin{eqnarray}
\label{SuppressFact}
{\mathcal A}_1 &=& \frac{[\kappa_2 s_{2\beta}+(\xi v)c_{2\beta}]}{m_\rho^2}\frac{(\Delta m^2_+)^2 \Delta m_0^2}{m_\rho^2 v^2},\nonumber\\
{\mathcal A}_2 &=& \frac{ \xi v}{m^2_\rho}\frac{\Delta m^2_+\Delta m_0^2}{ v^2},
\end{eqnarray}
where $m_\rho$ denotes the mass of the doubly charged scalar $\rho^{++}$, $x_i = m_i/v$ ($i=a,b$), $\Delta m_+^2 = m_{H_1^+}^2-m_{H_2^+}^2$ and $\Delta m_0^2=m_{H_0}^2-m_{A_0}^2$. 
 Note that the powers of $m_\rho$ in the denominators in Eq.~(\ref{SuppressFact}) are just to make the integrals ${\cal I}_{1,2}$ dimensionless for convenience, rather than their actual scaling dimensions.
The details of the derivations of Eqs.~(\ref{Mass}) and (\ref{SuppressFact}) as well as the precise definitions of the dimensionless integrals ${\cal I}_{1,2}$ are contained in Appendix~\ref{Details}. In our work, we have applied different widely-used softwares and packages to reliably perform the numerical integration of ${\cal I}_{1,2}$, such as {\sf Mathematica}, {\sf SecDec}~\cite{SecDec}, and {\sf GSL}~\cite{GSL}. 
 As a result, we find that the benchmark point given in the first version of Ref.~\cite{Gustafsson:2012vj} before its Erratum generically predicts the neutrino masses typically about two orders smaller than the measured ones, no matter what value of the coupling $\xi$ is if it is within the perturbative region $\xi \leq 5$~\cite{Nebot:2007bc}, which is also discussed in detail in Appendix~\ref{Details}.

It should be mentioned that the neutrino masses in Eq.~(\ref{Mass}) are proportional to $s_{2\beta}$. With our numerical studies, we conclude that the neutrino masses are usually insufficient to explain the oscillation data in most parameter spaces except those with large couplings $\kappa_2$ and $\xi$ and large mass splittings $\Delta m_+^2$ and $\Delta m_0^2$. In order not to introduce an extra suppression, we take the maximum value of $s_{2\beta}=1$, {\it i.e.}, $\beta=\pi/4$ in our following discussions.

\section{Neutrino Mass Matrix}~\label{NuPhys}
Currently, the mass differences and the mixings among three active neutrinos are measured to a very high precision, with the recent worldwide best-fit values as follows~\cite{PDG}:
\begin{eqnarray}\label{mass_measure}
&& \Delta m^2_{\rm sun} = (7.54^{+0.26}_{-0.22})\times 10^{-5} ~\mathrm{eV}^2,\quad \quad |\Delta m_{\rm atm}^2| = (2.43 ^{+0.06}_{-0.06})\times 10^{-3} ~\mathrm{eV}^2,\nonumber\\
&& \sin^2 \theta_{12} = 0.308\pm 0.017,\quad \sin^2 \theta_{23}=0.437^{+0.033}_{-0.023},\quad \sin^2 \theta_{13} = 0.0234^{+0.0020}_{-0.0019} .
\end{eqnarray}
The remaining questions are the pattern of the neutrino mass hierarchy and the four undetermined parameters:  the smallest neutrino mass $m_0$ and three CP violating phases, $\delta$ (Dirac) and $\alpha_{21,31}$ (Majorana), in the standard parametrization of the neutrino mixing matrix (see Ref.~\cite{PDG}).

To investigate the above questions in the context of the cocktail model, let us begin our discussion by noticing that the form of the neutrino mass matrix, that is, the relative size of each element, is determined by the Yukawa couplings $C_{ab}$. If all $C_{ab}$ are assumed to be of ${\mathcal O}(1)$, it is generically expected that the neutrino mass matrix should be in the form of the normal hierarchy since the mass elements are proportional to $x_a x_b$. And the elements $(m_\nu)_{ee,e\mu}$ should be much smaller than others due to the hierarchy $x_e \ll x_\mu \ll x_\tau$. With this expectation, we focus on the parameter space in which $(m_\nu)_{ee,e\mu}$ are approximately zero compared to other elements. This restriction amounts to four constraints to the active neutrino mass matrix, fixing the four known parameters to be
\begin{eqnarray}\label{solution}
m_0 = 5.14\times 10^{-3} {\mathrm eV} ,\quad \quad \delta = 1.89, \quad\quad \alpha_{21}=2.80, \quad\quad \alpha_{31} = 1.67 .
\end{eqnarray}
Consequently, the neutrino mass matrix can be predicted as:
\begin{eqnarray}\label{NuMass_bench}
m_\nu =\left(
      \begin{array}{ccc}
        \approx 0 & \approx 0 & 10.1 \\
        \approx 0 & -5.01 & 0.0980 \\
        10.1 & 0.0980 & -4.77
      \end{array}
    \right)\times 10^{-3} + i \left( \begin{array}{ccc}
   \approx 0 & \approx 0  &  0.23 \\
   \approx 0 & -2.37  & -2.33 \\
    0.23  & -2.33  & -2.74
    \end{array} \right) \times 10^{-2}
    ~\mathrm{eV},
\end{eqnarray}
where we have only used the central values in Eq.~(\ref{mass_measure}).
Note that $(m_\nu)_{ee,e\mu}$ are very sensitive to the choice of $m_0$ and the CP phases, 
so that if we keep $(m_\nu)_{ee,e\mu}$ small enough, the unknown parameters and the resulted neutrino masses cannot deviate the benchmark in Eqs.~(\ref{solution}) and (\ref{NuMass_bench}) much.

With  Eq.~(\ref{NuMass_bench}) and the formula for the cocktail model in Eq.~(\ref{Mass}), we can determine the Yukawa coupling matrix up to only one unknown parameter $|C_{e\tau}|$, given by
\begin{eqnarray}\label{YukMat}
C_{ab} = \left(
 \begin{array}{ccc}
 \leq {\cal O}(10^{-2}) & \leq {\cal O}(10^{-2})  & e^{0.224 i} \\
 \leq {\cal O}(10^{-2})  &  1.90 \times 10^{-1} e^{-1.78 i}   &  1.08\times 10^{-2} e^{-1.56 i}  \\
 e^{0.224 i}  &  1.08\times 10^{-2}e^{-1.56 i}   &  7.73\times 10^{-4}e^{-1.74 i}
 \end{array}
 \right)\times |C_{e\tau}|.
\end{eqnarray}
We will take advantage of this rigid structure of the Yukawa coupling matrix in our discussion of LFV processes by expressing their constraints in terms of $|C_{e\tau}|$. Note that the elements $C_{ee,e\mu}$ cannot be determined with the benchmark point in Eq.~(\ref{NuMass_bench}). Actually, they can be tuned to be as small as possible without affecting the neutrino mass matrix form. The largest orders shown in Eq.~(\ref{YukMat}) are obtained by combining the constraints from the  $0\nu\beta\beta$ decay and LFV processes.

\begin{figure}[ht]
\centering
\includegraphics[width=5cm]{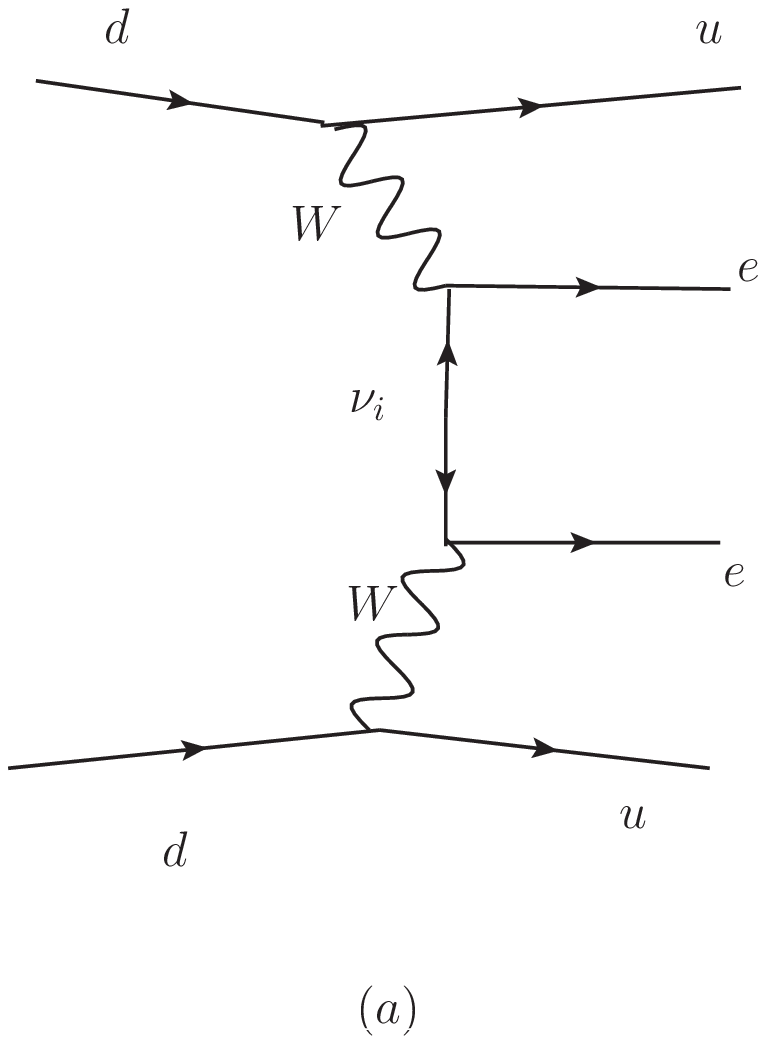}
\includegraphics[width=7cm]{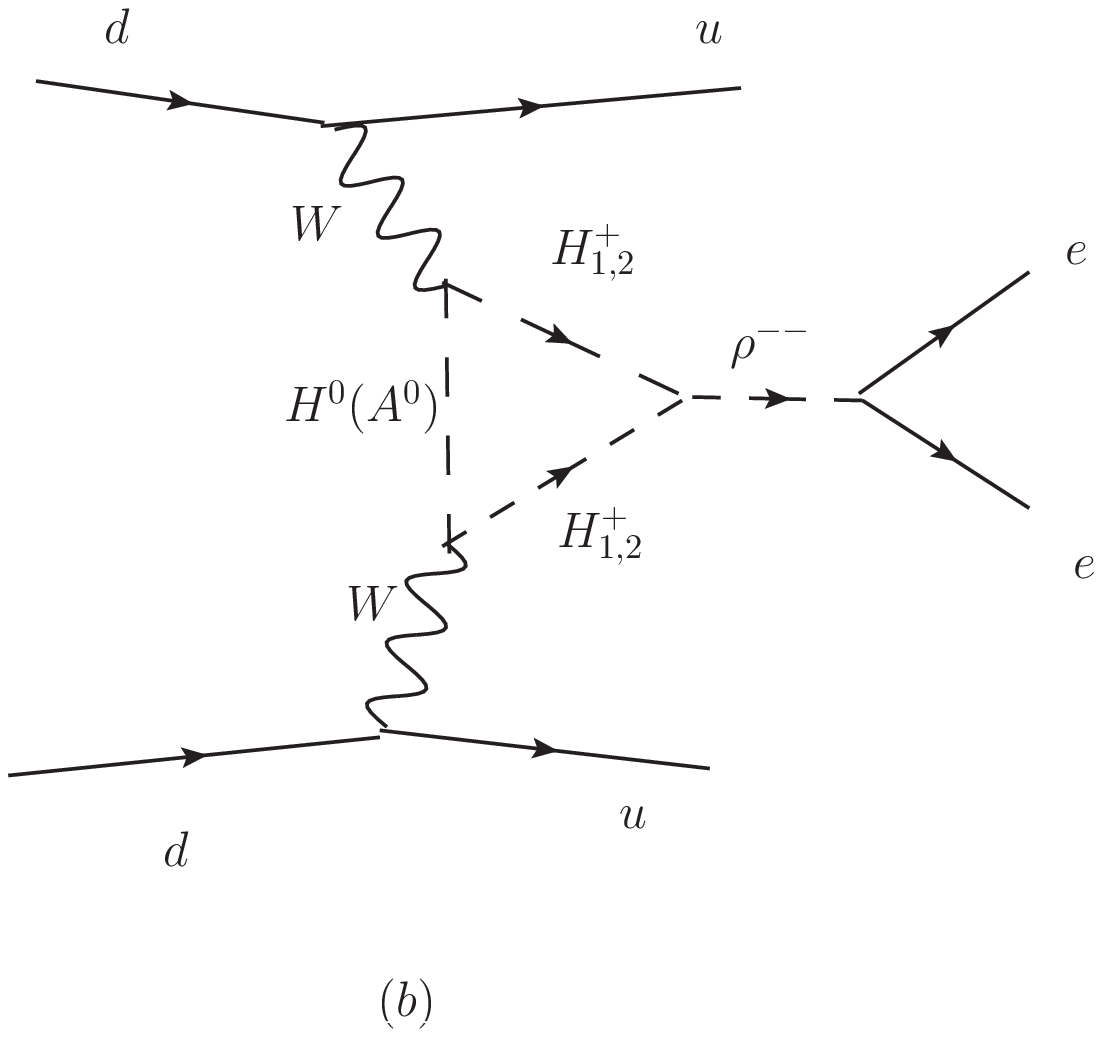}
\caption{$0\nu\beta\beta$-decay from (a) long-distance and (b) short-distance diagrams.}
\label{Fig_0nbb}
\end{figure}

The neutrinoless double beta decay, as a lepton number violating process, should exist with a non-zero $(m_{\nu})_{ee}$, which is equivalent to a non-zero $C_{ee}$ in the cocktail model. In the conventional neutrino mass generation models, such as the type-II seesaw model~\cite{TypeII-Seesaw}, the long-distance contribution as shown in Fig.~\ref{Fig_0nbb}(a) dominates the decay process, while for the cocktail model models which can generate the effective coupling $\rho^{--} W_\mu^{+} W^{\mu +}$, the short-distance channel shown in Fig.~\ref{Fig_0nbb}b gives the contribution several orders larger than the long-distance one~\cite{Pas:2000vn,Deppisch:2012nb,Chen:2006vn,Gustafsson:2014vpa,King:2014uha}.
This feature can be traced to the fact that the amplitude of Fig.~\ref{Fig_0nbb}b is proportional to $C_{ee}$, rather than $m_{ee}$, which is further suppressed by the small electron mass squared $m_e^2$. In this way, some parameter spaces have already been probed and constrained by the current $0\nu\beta\beta$ experiments. Since the energy transfer in the $0\nu\beta\beta$ decay is only of order $100$ MeV, the  short-distance contribution in Fig.~\ref{Fig_0nbb}b to the half-life for the $0\nu\beta\beta$ decay in the cocktail model can be expressed by~\cite{Gustafsson:2014vpa}
\begin{eqnarray}
T_{1/2}^{\nu0\beta\beta}=\Big[4m_p^2G_{01}|\mathcal{A}|^2|\mathcal{M}_{3}|^2\Big]^{-1}\;,\label{Eq_hafelife}
\end{eqnarray}
where $m_p$ is the mass of the proton, $G_{01}$ the phase space factor,
\begin{eqnarray}
{\mathcal A}&=&\frac{C_{ee}s_{2\beta}\Delta m_+^2}{8\pi^2 m_\rho^2}\Big\{
[\kappa_2 \Delta m_+^2 s_{2\beta}-\xi v (c_\beta^2 m_{H_2}^2+s_\beta^2 m_{H_1}^2)][F_{H_1^+,H_2^+,H_0}-F_{H_1^+,H_2^+,H_0}]\nonumber\\
&& -\xi v [m_{H_0}^2F_{H_1^+,H_2^+,A_0}-m_{A_0}^2F_{H_1^+,H_2^+,A_0}]
\Big\}\;,
\end{eqnarray}
with 
\begin{eqnarray}
F_{a,b,c}=\int_0^1 du \int_0^1 dv \frac{u^3v(1-v)}{[uv m_a^2+u(1-v) m_b^2 +(1-u-v)m_c^2]^2}\;,
\end{eqnarray}
and $\mathcal{M}_{3}$ the nuclear matrix element enveloping the operator $\bar u_L\gamma^\mu d_L \bar u_L\gamma_\mu d_L \bar e_Re_R^c$, as defined in Refs.~\cite{Doi:1985dx,Suhonen:1998ck, Deppisch:2012nb}.  The numerical values of $G_{01}$ and ${\cal M}_3$ for several conventional targets are collected in Table~\ref{Tab_NME}~\cite{Deppisch:2012nb}.
\begin{table}[ht]
\caption{$G_{01}$ (in unit $10^{-14}{\rm yr}^{-1}$) and $|\mathcal{M}_3|$ for different nuclei, where the numerical values are taken from Ref.~\cite{Deppisch:2012nb}.}
\begin{tabular}{ccccccc}
\hline
&$^{76}$Ge&$^{136}$Xe&$^{150}$Nd&$^{130}$Te&$^{82}$Se&$^{100}$Mo\\
\hline
$G_{01}$&0.640&4.73&21.0&4.44&2.82&4.58\\
\hline
$|\mathcal{M}_3|$&209&107&305&193&188&241\\
\hline
\end{tabular}
\label{Tab_NME}
\end{table}

For a rough estimation, we take a benchmark point for the scalar masses and related coupling constants as an illustration, given by 
$\kappa_2=6\,{\rm TeV}$, $\xi=5$, $m_{H_1}=200\,{\rm GeV}$, $m_{H_2}=720\,{\rm GeV}$, $m_{H_0}=70\,{\rm GeV}$, $m_{A_0}=430\,{\rm GeV}$, and $m_{\rho}=2\,{\rm TeV}$. From the current experimental detections~\cite{Agostini:2013mzu, KamLANDZen, Argyriades:2008pr, Arnaboldi:2008ds, Arnold:2005rz, Barabash:2010bd}, we can obtain the upper bound on $C_{ee}$ by applying Eq.~(\ref{Eq_hafelife}), with the results listed in Table.~\ref{Tab_maxCee}.
\begin{table}[ht]
\caption{Experimental lower bounds on the half-life of $0\nu\beta\beta$ with the corresponding maximal values of $|C_{ee}|$. 
}
\begin{tabular}{lcl}
\hline
&$>T_{\rm exp}(10^{25}{\rm yr})$\;\;\;\;&$|C_{ee}|_{\rm max}$\\
\hline
GERDA-1($^{76}$Ge)~\cite{Agostini:2013mzu}&2.1&0.0015\\
KamLAND-Zen($^{136}$Xe)~\cite{KamLANDZen}&1.9&0.0011\\
NEMO-3($^{150}$Nd)~\cite{Argyriades:2008pr}&0.0018&0.0060\\
CUORICINO($^{130}$Te)~\cite{Arnaboldi:2008ds}&0.3&0.0016\\
NEMO-3($^{82}$Se)~\cite{Arnold:2005rz,Barabash:2010bd}&0.036&0.0059\\
NEMO-3($^{100}$Mo)~\cite{Barabash:2010bd}&0.11&0.0021\\
\hline
\end{tabular}
\label{Tab_maxCee}
\end{table}
Generically, $C_{ee}$ should be less than $10^{-3}$ to fulfill all the present experimental constraints with the most tightly constraints on $C_{ee}$ from the detections for the targets $^{76}$Ge and $^{136}$Xe. On the other hand, models with the long-distance dominance usually predict an undetectable half life for the $0\nu\beta\beta$ decay~\cite{Gustafsson:2014vpa}. Therefore, future experiments with a higher sensitivity~\cite{future} could help to distinguish the cocktail model from the conventional ones.
\section{Flavor Constraints}\label{flavor}
The overall size of the Yukawa coupling matrix in Eq.~(\ref{YukMat}) is mostly constrained by the LFV processes mediated by the doubly-charged scalar $\rho^{++}$, in which the most relevant ones can be categorized into two kinds: $\ell_0^\mp \rightarrow \ell_1^\pm \ell_2^\mp \ell_3^\mp$ and $\ell_0^\pm \rightarrow \ell_1^\pm \gamma$, and the corresponding formulae are listed as
\begin{eqnarray}\label{LFV}
&& {\mathcal B}(\ell_0^\mp \rightarrow \ell_1^\pm \ell_2^\mp \ell_3^\mp) = \frac{|C_{\ell_1 \ell_0} C_{\ell_2\ell_3}^*|^2}{2 m_\rho^4 G_F^2} \frac{m_{\ell_0}^5}{m_{\mu}^5} {\mathcal B}(\mu^- \rightarrow e^- \overline{\nu}_e \nu_\mu),\nonumber \\
&& {\mathcal B}(\ell^\pm_0 \rightarrow \ell^\pm_1 \gamma) = \frac{\alpha_{\rm em}}{3\pi} \frac{|\sum_\ell C^*_{\ell_1\ell} C_{\ell\ell_0} |^2}{G_F^2 m_\rho^4} \frac{m_{\ell_0}^5}{m_{\mu}^5} {\mathcal B}(\mu^- \rightarrow e^- \overline{\nu}_e \nu_\mu).
\end{eqnarray}
With Eq.~(\ref{LFV}), the bounds on the various LFV processes~\cite{Adam:2013mnn,Hayasaka:2010np,PDG} can be translated into the ones on the Yukawa couplings:
\begin{eqnarray}
&&{\mathcal B}(\mu^+ \rightarrow e^+ \gamma) < 5.7\times 10^{-13}:\quad\quad |\sum_\ell C_{\ell \mu}C^*_{\ell e}|< 3.16\times 10^{-4} (m_\rho/{\rm TeV})^2,  \nonumber\\
&&{\mathcal B}(\mu^- \to 3e) < 1.0 \times 10^{-12}:\quad\quad |C_{e\mu}C^*_{ee}|< 2.33\times 10^{-5} (m_\rho/{\rm TeV})^2, \nonumber\\
&&{\mathcal B}(\tau^- \rightarrow 3e) < 2.7\times 10^{-8}:\quad\quad |C_{e\tau}C^*_{ee}|< 9.1\times 10^{-3} (m_\rho/{\rm TeV})^2, \nonumber\\
&&{\mathcal B}(\tau^-\rightarrow 3\mu) < 2.1\times 10^{-8}:\quad\quad |C_{\mu\tau}C^*_{\mu\mu}|< 8.0\times 10^{-3} (m_\rho/{\rm TeV})^2, \nonumber\\
&&{\mathcal B}(\tau^-\rightarrow e^-\mu^+\mu^-) < 2.7\times 10^{-8}:\quad\quad |C_{\mu\tau}C^*_{e\mu}|< 6.42 \times 10^{-3} (m_\rho/{\rm TeV})^2, \nonumber\\
&&{\mathcal B}(\tau^-\rightarrow \mu^- e^+ e^-)< 1.8\times 10^{-8}:\quad\quad |C_{e\tau}C^*_{e\mu}|< 5.24 \times 10^{-3} (m_\rho/{\rm TeV})^2, \nonumber\\
&&{\mathcal B}(\tau^-\rightarrow e^+\mu^-\mu^-) < 1.7\times 10^{-8}:\quad\quad |C_{e\tau}C^*_{\mu\mu}|< 7.21 \times 10^{-3} (m_\rho/{\rm TeV})^2, \nonumber\\
&&{\mathcal B}(\tau^-\rightarrow \mu^+ e^- e^-) < 1.5 \times 10^{-8}:\quad\quad |C_{\mu\tau}C^*_{ee}|< 6.77 \times 10^{-3} (m_\rho/{\rm TeV})^2.
\end{eqnarray}
These constraints, together with the typical Yukawa matrix pattern in Eq.~(\ref{YukMat}), can yield the upper bound on the overall size of the Yukawa coupling $|C_{e\tau}|< 0.168 (m_\rho/{\rm TeV})$, which comes mainly from the process $\mu \rightarrow e \gamma$. For the later convenience, we shall take $|C_{e\tau}|=0.15(m_\rho/{\rm TeV})$ in our numerical exploration of the parameter space. For $C_{ee,e\mu}$, the bound on the branching ratio of $\mu \to 3 e$ restricts $C_{ee,e\mu} \leq {\cal O}(10^{-3})$, which are consistent with the aforementioned $0\nu\beta\beta$ decay constraints.

\section{Electroweak Precision Test Constraints}\label{EWPT}
Since all of the newly introduced particles carry EW charges, the cocktail model is also well constrained by the EWPTs at the LEP, especially the $T$ parameter~\cite{Grimus:2008nb,Baak:2012kk}, for which the new one-loop correction is as follows~\cite{Gustafsson:2012vj}:
\begin{eqnarray}\label{Eq_EWPT}
\Delta T &=& \frac{1}{16\pi m_W^2 s_W^2} \Big[c_\beta^2 \left( F_{H^+_1, H_0} + F_{H_1^+, A_0} \right) +s^2_\beta \left( F_{H_2^+, H_0} + F_{H_2^+, A_0} \right) \nonumber\\
&& -2 c^2_\beta s_\beta^2 F_{H_1^+, H_2^+} -F_{H_0, A_0} \Big],
\end{eqnarray}
where
\begin{eqnarray}
F_{i,j} = \frac{m_i^2+m_j^2}{2}-\frac{m_i^2 m_j^2}{m_i^2-m_j^2}\ln\frac{m_i^2}{m_j^2},
\end{eqnarray}
and $s_W$($c_W$) is the sine(cosine) of the Weinberg angle $\theta_W$. As already pointed in Ref.~\cite{Gustafsson:2012vj}, 
the cancelation between the charged and neutral states becomes possible, resulting in an extended parameter space. Especially, the present model allows a large mass splitting between the two neutral (charged) particles.

\section{Dark Matter Physics and Collider Constraints}\label{DM-collider}
DM physics and the collider searches have already provided interesting constraints on the cocktail model. Since the cocktail model is very similar to the widely-studied inert doublet model (IDM)~\cite{IDM} in the $Z_2$-odd sector except for the additional singly charged scalar, we would expect that the results about the DM properties in the IDM could be applied directly. In the following, we just summarize some of relevant conclusions from the most recent global fitting studies in Ref.~\cite{Arhrib:2013ela}. Other aspects of the IDM can be referred to Refs.~\cite{Arhrib:2013ela,Gustafsson:2012aj,Hambye:2009pw,Belanger:2013xza}.

In the cocktail model, there is no preference of the neutral scalar $H_0$ or the pseudoscalar $A_0$ to be the dark matter candidate. Thus, without loss of generality, we assume that the lightest $Z_2$-odd particle is $H_0$. According to the analysis in Ref.~\cite{Arhrib:2013ela}, there are two regions, $60~ {\rm GeV} < m_{H_0} < 75~ {\rm GeV}$ (low mass) and $m_{H_0} > 500$~GeV (high mass), that can give rise to the correct relic DM density while satisfying all other experimental constraints, including LHC searches, direct detection bounds from LUX and XENON100, and indirect signals from AMS-02 and Fermi-LAT, with the low mass region favored by the fit. In addition, in the large mass region, it is crucial that $m_{H_0} \approx m_{A_0}$, which is required by the coannihilation of the DM $H_0$ with $A_0$ to generate the correct DM relics. On the other hand, it is clear from Eqs.~(\ref{Mass}) and (\ref{SuppressFact}) that the right amount of the neutrino masses needs a large enough mass difference between $H_0$ and $A_0$. Therefore, there is some tension between the DM relic density and the neutrino masses in the high mass region. In the following, we only focus on the low mass region and take $m_{H_0} = 70$~GeV as our benchmark point, which is also the best fitting point in Ref.~\cite{Arhrib:2013ela}. In this region, the right relic density in the Universe~\cite{Planck} can be obtained by the combination of three effects~\cite{Arhrib:2013ela,Gustafsson:2012aj}: the coannihilation among $H_0$, $A_0$ and $H_1^{\pm}$, the SM Higgs resonance enhancement, and the opening of the $W^+ W^-$ annihilation channel. Furthermore, if we restrict the coupling $-\lambda_5 (\Phi_1^\dagger \Phi_2)^2/2$ to be within the perturbative region with $|\lambda_5| <5$, we find that the upper bound for the pseudoscalar mass is $m_{H_0}<m_{A_0} < 555$~GeV. However, the LEP has excluded models with $m_{A_0} \leq 100$~GeV when $m_{H_0}=70$~GeV~\cite{Lundstrom:2008ai}.

The allowed range of the masses for the charged particles are well constrained by the EWPTs, especially the $T$ parameter. Due to the mixing 
involving with  the extra $SU(2)_L$ singlet charged scalar, it is clear that the results in the present cocktail model vastly differ from those in the IDM as shown in Eq.~(\ref{Eq_EWPT}), so that we cannot directly copy the conclusion in Ref.~\cite{Arhrib:2013ela} here. Rather, if we require the heavier singly charged scalar $H_2^{+}$ to be less than 1~TeV and the mass splitting $\Delta m_+^2$ large enough to generate measured values of the neutrino masses, $m_{H_1^+}$ should not exceed 500~GeV from our numerical studies. Thus, we take three benchmark points with $m_{H_1^+}=90$, 200, and 300 GeV, respectively, which are all allowed by the LEP constraints $m_{H_1^+}\leq 70 -90$~GeV~\cite{Pierce:2007ut}. Note that the latest 8~TeV ATLAS~\cite{ATLAS8_Chargino1,ATLAS8_Chargino2,ATLAS8_Chargino3} and CMS~\cite{CMS8_Chargino1,CMS8_Chargino2} bounds on the chargino and neutralino masses cannot be applied here, since either they assumed the equal mass of the lightest chargino and second-lightest neutralino in the associated production channel~\cite{ATLAS8_Chargino1,ATLAS8_Chargino2,CMS8_Chargino1,CMS8_Chargino2} or the constraining power on the lightest chargino mass was only confined within the DM masses smaller than about 30~GeV in the chargino pair production one~\cite{ATLAS8_Chargino2}.
For the doubly charged scalar $\rho^{++}$, the most stringent bounds on its mass are 409~GeV and 459~GeV for the ATLAS~\cite{Rho_ATLAS7} and CMS~\cite{Rho_CMS7} 7~TeV data, respectively.


\section{Numerical Results}\label{NumericalRes}
\begin{figure}[ht]
\centering
\includegraphics[angle=-90,width=0.45\textwidth]{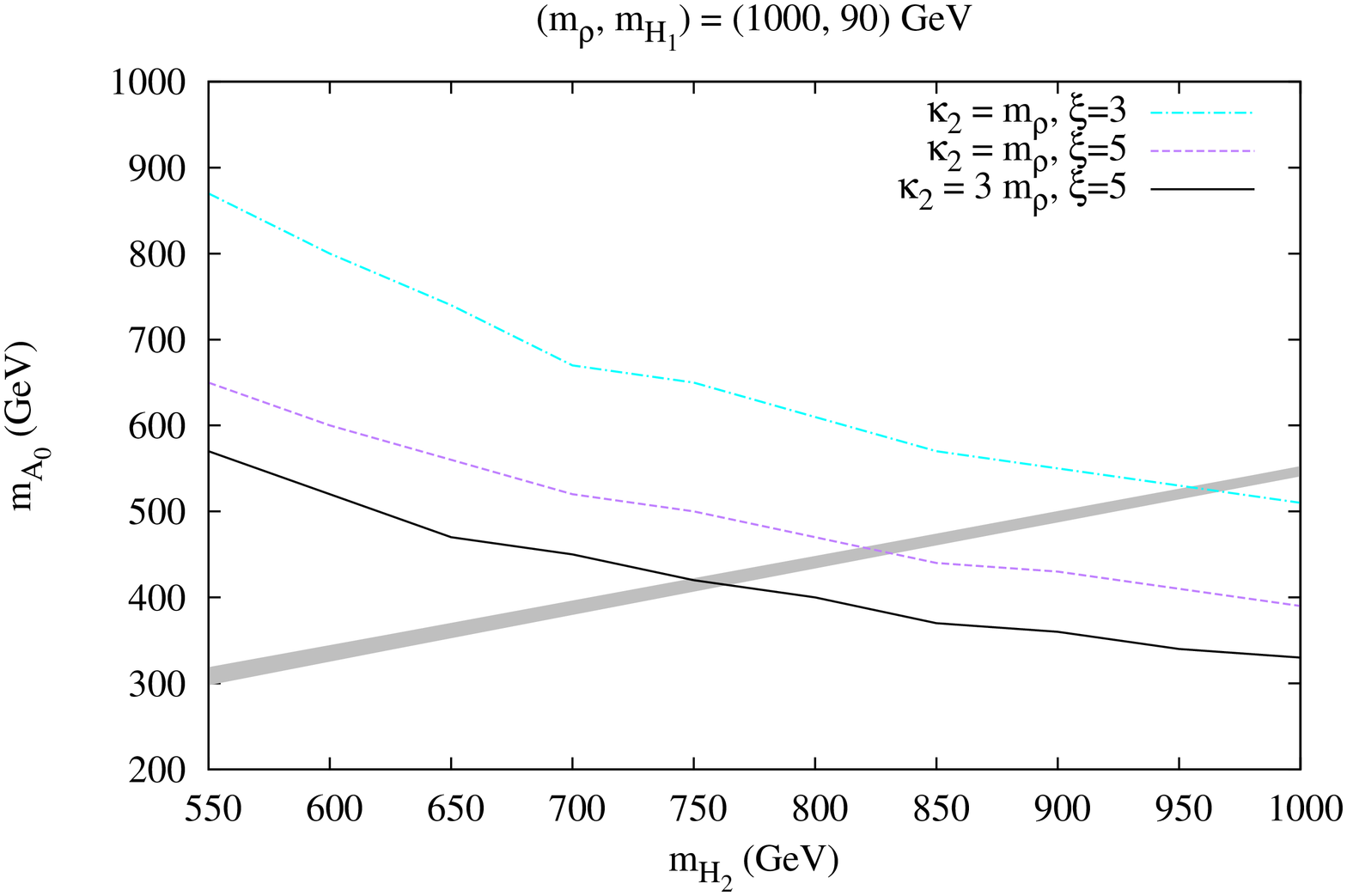}
\includegraphics[angle=-90,width=0.45\textwidth]{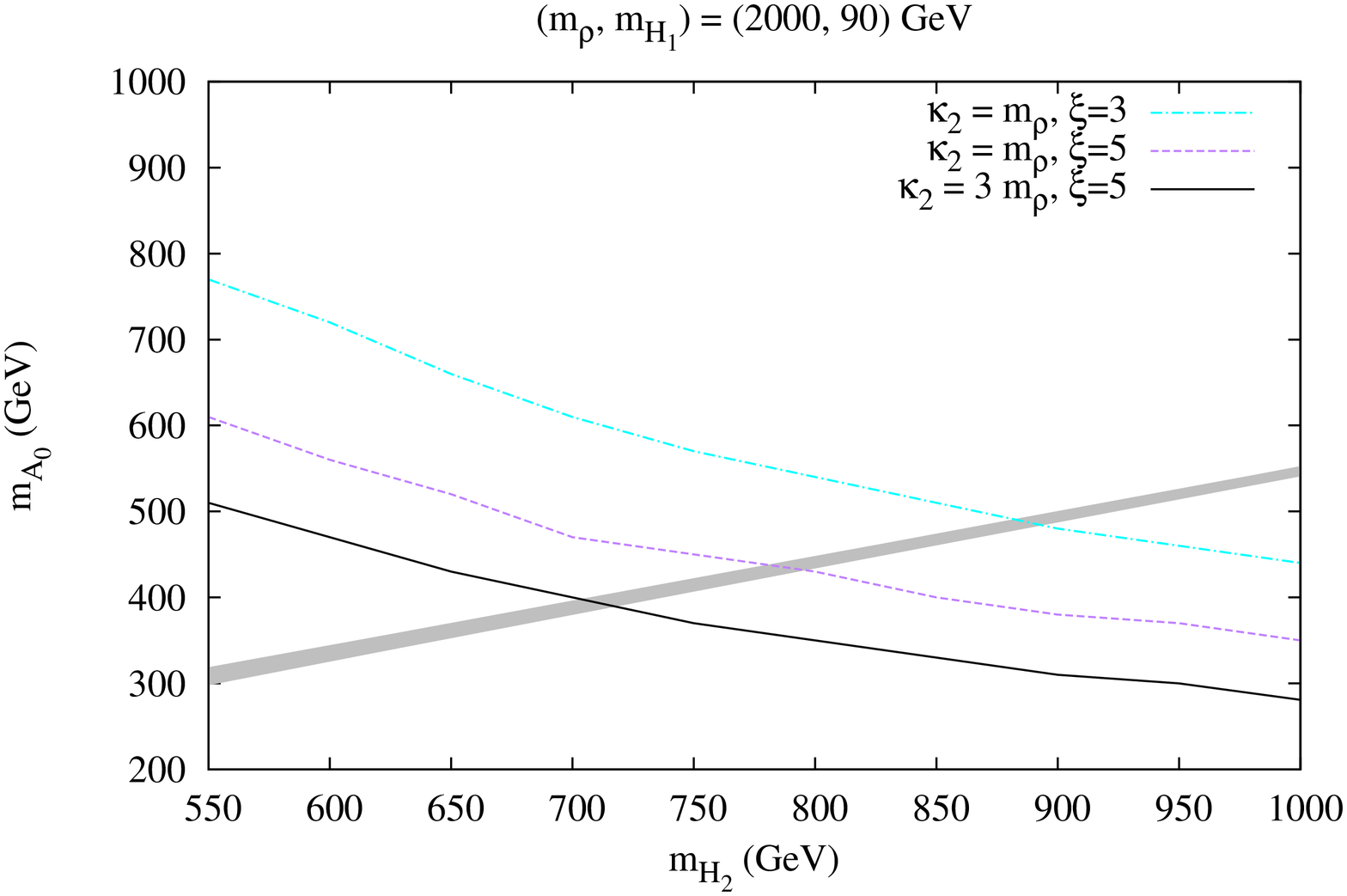}
\includegraphics[angle=-90,width=0.45\textwidth]{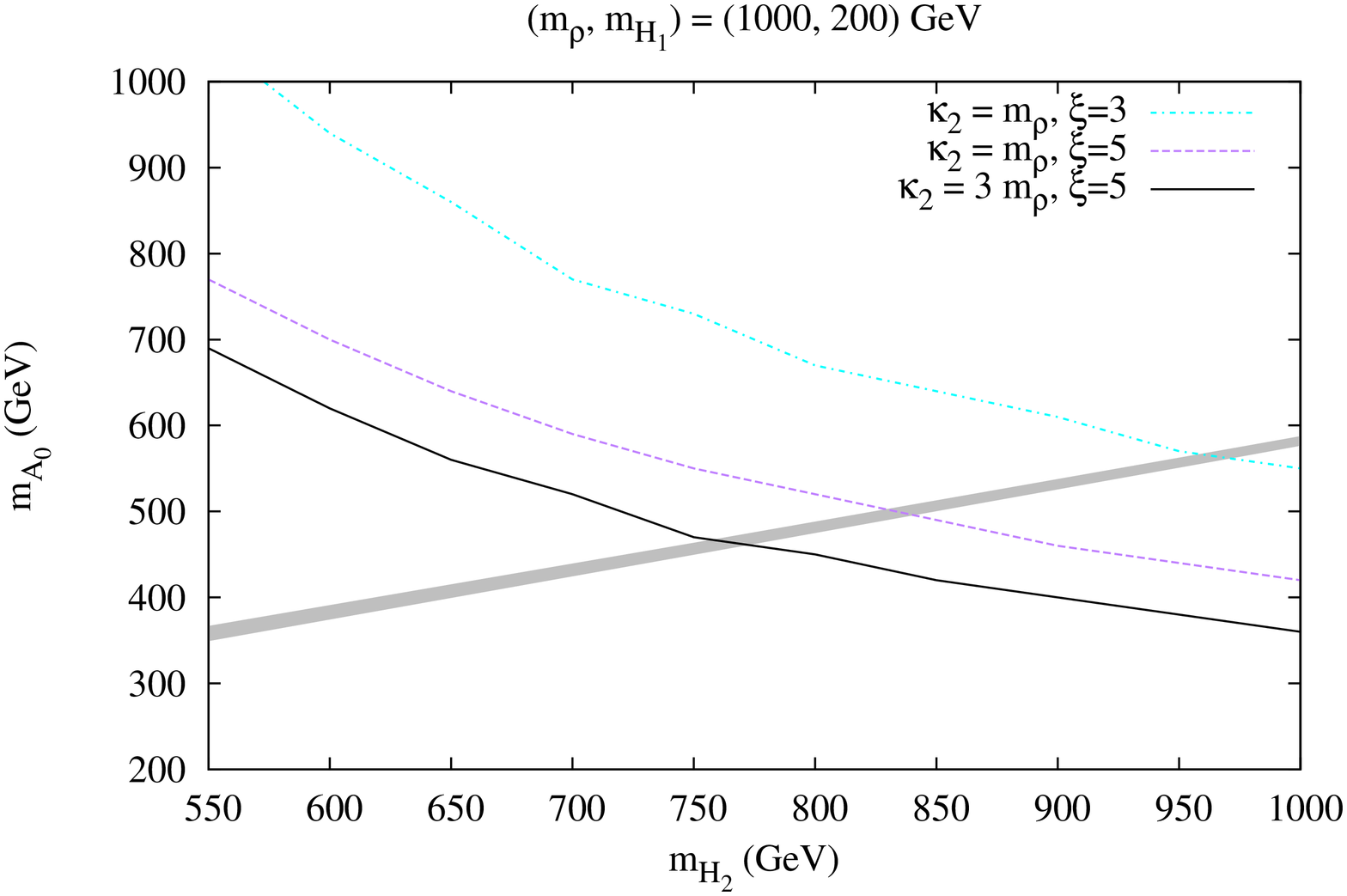}
\includegraphics[angle=-90,width=0.45\textwidth]{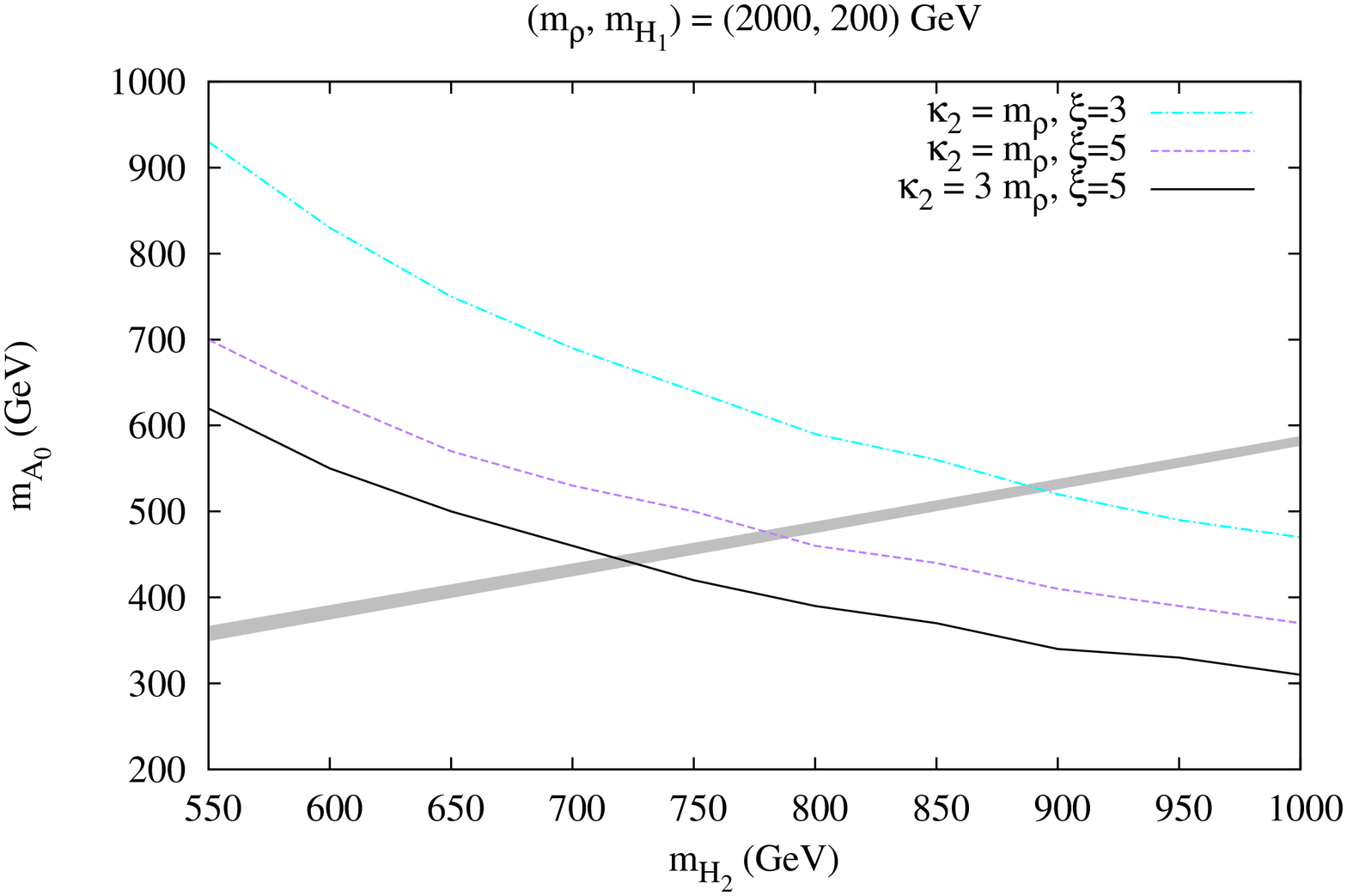}
\includegraphics[angle=-90,width=0.45\textwidth]{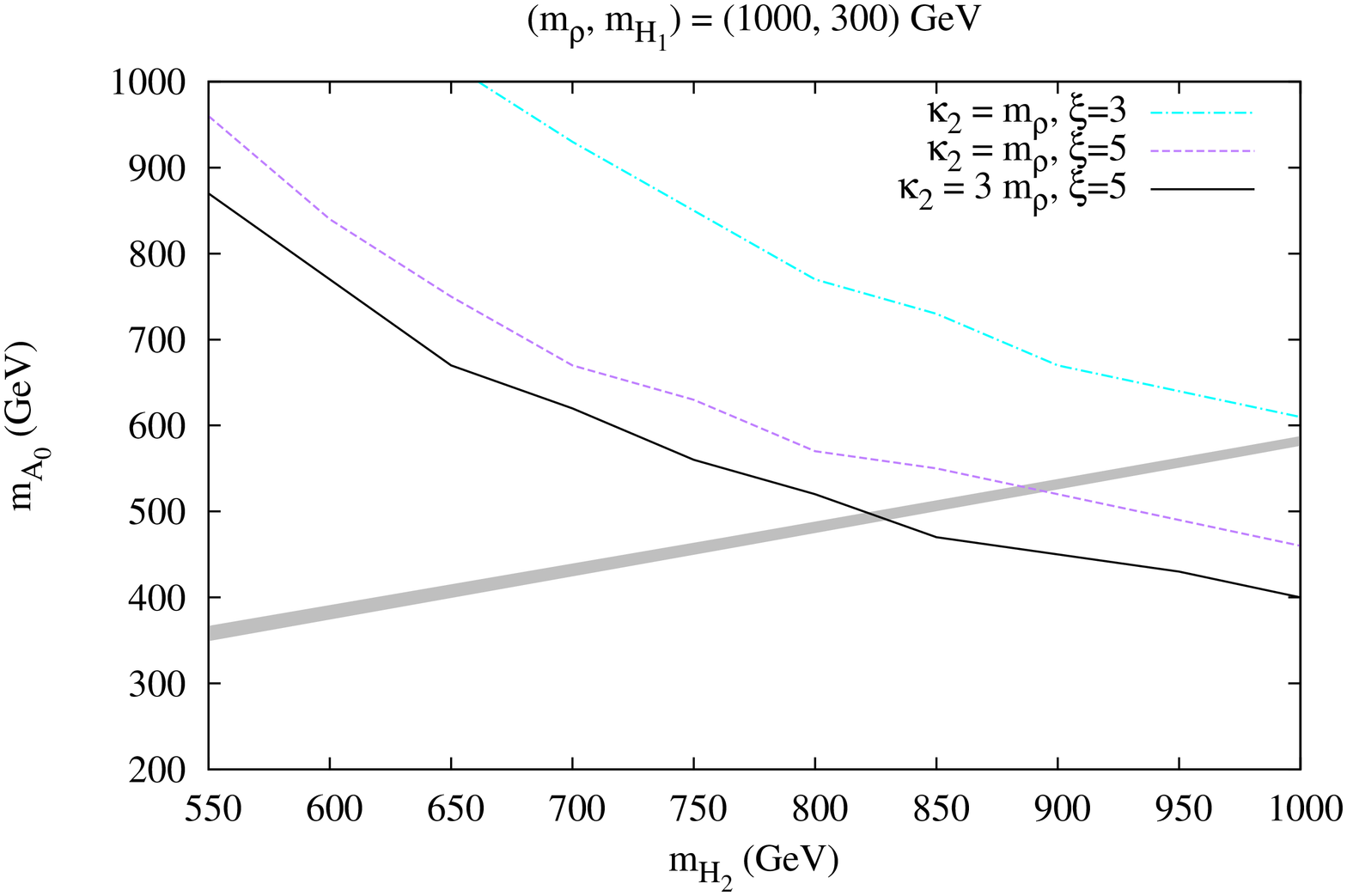}
\includegraphics[angle=-90,width=0.45\textwidth]{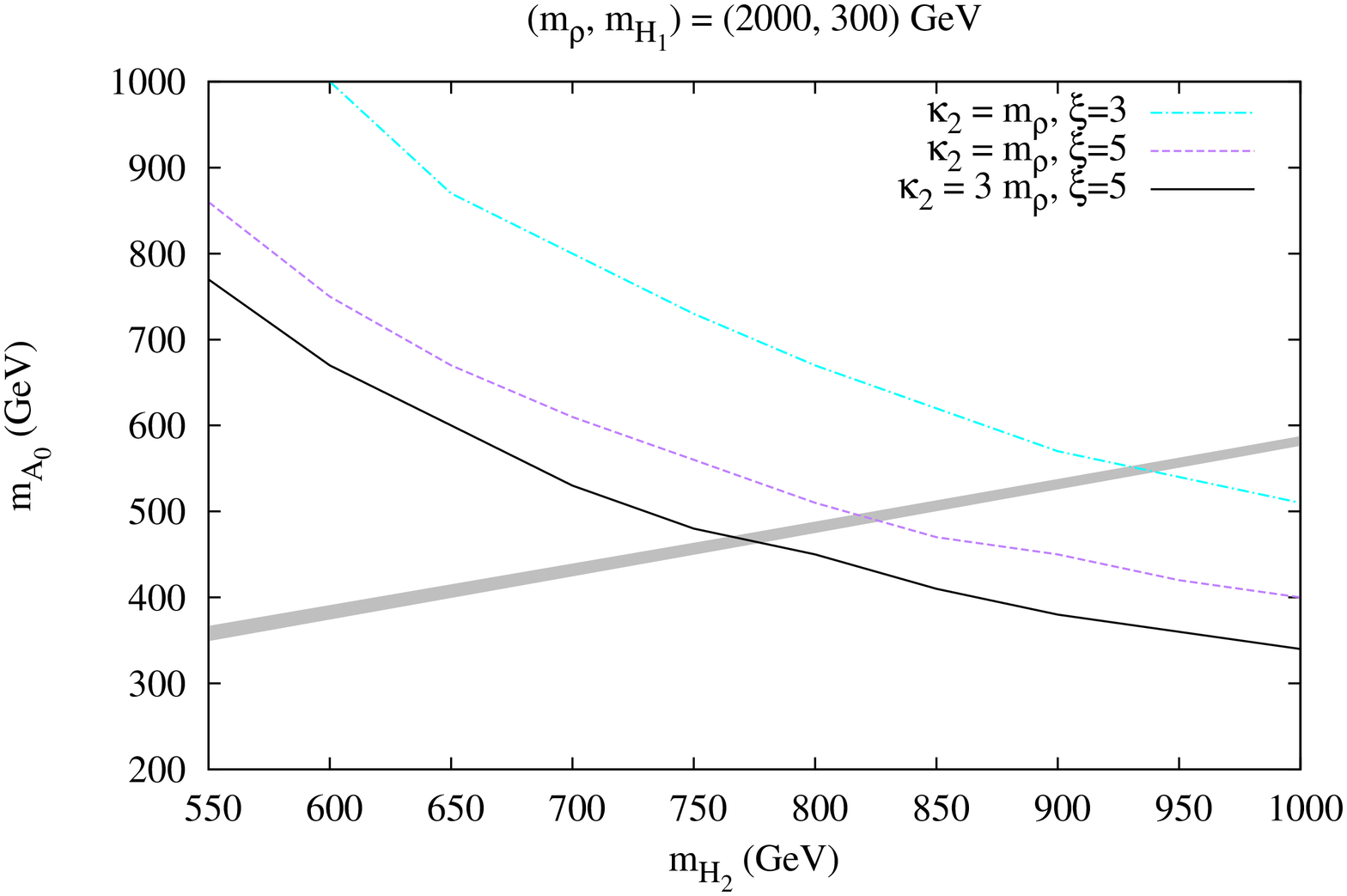}
\caption{Benchmark points for allowed parameter space, where the grey bands represent  1$\sigma$ errors
of the allowed parameter spaces from $\Delta T$.}
\label{Result}
\end{figure}

Instead of the exploration of the whole parameter space, we only present some benchmark points of phenomenological interests. In particular, we focus on the particle spectra in which $m_\rho = 1$ and 2~TeV with $m_{H_1}=90$, 200, and 300~GeV. If we further confine the couplings $(\kappa_2,\xi)$ within the perturbative region and take the following characteristic values, such as (0.7$m_\rho$, 0), ($m_\rho$,0), ($m_\rho$, 3), ($m_\rho$, 5) and ($3 m_\rho$, 5), the allowed parameter space which can give the correct size of the neutrino masses while satisfying flavor constraints are plotted as lines in the $m_{H_2}$-$m_{A_0}$ plane in Fig.~\ref{Result}, which can be compared with the allowed parameter spaces (grey bands) from $\Delta T$ with the
1$\sigma$ errors.

The general feature as seen from the diagrams in Fig.~\ref{Result} is that  with the relatively large couplings $\kappa_2 \gtrsim m_\rho$
 and $\xi \gtrsim 3$, it is easy to obtain the correct neutrino masses while satisfying all constraints. And the final results are more sensitive to $\xi$ than $\kappa_2$, since the integral ${\cal I}_2$ is generically larger than ${\cal I}_1$ with the chosen mass spectra. In particular, when $(\kappa_2,\xi)$=(0.7$m_\rho$, 0) and ($m_\rho$,0), we cannot find any solution of $(m_{H_2},m_{A_0})$ to realize enough neutrino masses in our phenomenologically interesting region, so we do not plot any lines for these two benchmarks in Fig.~\ref{Result}. 
Moreover, by comparing two diagrams in each line, the increase of $m_\rho$ allows more parameter space in the $m_{H_2}$-$m_{A_0}$ plane, while the careful examination of the three diagrams in either column shows that the parameter space shrinks when we enlarge $m_{H_1}$. The former phenomenon can be attributed to our assumption of $|C_{e\tau}|=0.15(m_\rho/{\rm TeV})$ which effectively makes larger Yukawa couplings when amplifying $m_\rho$, while the latter can be understood as the decrease of the neutral scalar mass difference when $m_{H_1}$ increases.

\section{Conclusions}\label{Conclusion}
We have explored the cocktail model introduced in Ref.~\cite{Gustafsson:2012vj}, which is interesting because it provides a connection between the origin of the small neutrino masses and the dark matter physics. In particular, we have shown 
the detailed derivation of the neutrino mass formulae in Eq.~(\ref{Mass}) from the three-loop cocktail diagrams, for which the subsequent loop integrals are calculated with the reliable numerical methods. Based on Eq.~(\ref{Mass}), the neutrino mass matrix is naturally predicted to be of the normal hierarchy type, with the nearly vanishing elements $(m_\nu)_{ee,e\mu}$. Consequently, the current data on the neutrino mass differences and mixings already fix the mass matrix to a high precision. By further considering the stringent constraints from the neutrinoless double beta decay, the low-energy LFV processes, the EWPT measurement of the $T$ parameter, the DM relic density, and the collider searches, the DM mass is confined in the narrow range $60~{\rm GeV} < m_{H_0} < 75~{\rm GeV}$, and the right order of the 
neutrino masses can only be obtained by the large mass splittings for the neutral and charged scalars as well as the large couplings of
 $\kappa_2 \sim m_\rho$ and $\xi\gtrsim 3$. It is interesting to point out that the $0\nu\beta\beta$ decay is predominately via the new short-distance contribution, which is typically larger than the usual long-distance one and has the possibility to be observed in the next-generation experiments.  


\begin{acknowledgments}
The work was supported in part by National Center for Theoretical Sciences, National Science
Council (NSC-101-2112-M-007-006-MY3) and National Tsing Hua
University (103N2724E1).
\end{acknowledgments}

\appendix
\section{Detailed Calculations of the Cocktail Diagrams }\label{Details}
In this appendix we compute the cocktail diagrams in the unitary gauge.
Without loss of generality, we separate the neutrino mass into two parts, one proportional to $\kappa_2$ and the other  $\xi v$, which will be calculated in the following two subsections. 
In our numerical calculation, we study both zero and nonzero cases for $\xi$.

\subsection{Integrals proportional to $\kappa_2$}
In the unitary gauge, there are only 8 Feynman diagrams in Fig.~\ref{FigCocktail}. Let us first focus on the upper triangle loops in the top of the diagrams, which are the only differences among the diagrams. The relevant pieces of the Lagrangian are:
\begin{eqnarray}\label{Lag_k}
\mathcal{L}&=&-\frac{ig_2}{\sqrt{2}}\Lambda^-W^{+\mu}\partial_\mu\Phi^{0}-\kappa_2\rho^{++}S^-S^-+{\rm h.c.}\nonumber\\
& = &-\frac{ig_2}{2}(c_\beta H_1^--s_\beta H_2^-)W^{+\mu}\partial_\mu{(H_0+iA_0)}\nonumber\\
&& - \kappa_2\rho^{++}(s_\beta^2 H_1^-H_1^-+2c_\beta s_\beta H_1^-H_2^-+c_\beta^2 H_2^-H_2^-)+\mathrm{h.c.}
\end{eqnarray}
where $\Lambda^+=c_\beta H_1^+-s_\beta H_2^+$ and $S^+=s_\beta H_1^++c_\beta H_2^+$.
We choose $\langle H\rangle=(0,\;v)^T$ with $v=173$~GeV. Using the Feynman rules, the triangle-loop factors involving $H_0$ are
\begin{eqnarray}
(H_0H_1H_1)&:& (\kappa_2)\frac{g_2^2 s^2_{2\beta}}{8} \frac{(2k-k_2)_\nu (-2k-k_1)_\mu}{(k^2-m_{H_0}^2)[(k+k_1)^2-m_{H_1}^2][(k-k_2)^2-m_{H_1}^2]}\;,\nonumber\\
(H_0H_1H_2)&:&-(\kappa_2)\frac{g_2^2 s^2_{2\beta}}{8} \frac{(2k-k_2)_\nu (-2k-k_1)_\mu}{(k^2-m_{H_0}^2)[(k+k_1)^2-m_{H_1}^2][(k-k_2)^2-m_{H_2}^2]}\;,\nonumber\\
(H_0H_2H_1)&:&-(\kappa_2)\frac{g_2^2 s^2_{2\beta}}{8} \frac{(2k-k_2)_\nu (-2k-k_1)_\mu}{(k^2-m_{H_0}^2)[(k+k_1)^2-m_{H_2}^2][(k-k_2)^2-m_{H_1}^2]}\;,\nonumber\\
(H_0H_2H_2)&:&(\kappa_2)\frac{g_2^2 s^2_{2\beta}}{8} \frac{(2k-k_2)_\nu (-2k-k_1)_\mu}{(k^2-m_{H_0}^2)[(k+k_1)^2-m_{H_2}^2][(k-k_2)^2-m_{H_2}^2]}\;,
\end{eqnarray}
while the corresponding factors for the diagrams involving the pseudoscalar $A_0$ are essentially the same with an additional minus sign.
Thus, by summing these 8 diagrams, we obtain
\begin{eqnarray}\label{TriangleSum_k}
&& (\kappa_2)\frac{g_2^2 s^2_{2\beta}}{8}\int\frac{d^4 k}{(2\pi)^4}\{(2k-k_2)_\nu (-2k-k_1)_\mu (\Delta m_+^2)^2 \Delta m_0^2\}/\{(k^2-m_{H_0}^2)(k^2-m_{A_0}^2) \nonumber\\
&& [(k+k_1)^2-m_{H_1}^2][(k+k_1)^2-m_{H_2}^2][(k-k_2)^2-m_{H_1}^2][(k-k_2)^2-m_{H_2}^2]\}\,,
\end{eqnarray}
where $\Delta m_+^2 = m_{H_1}^2-m_{H_2}^2$ and $\Delta m_0^2 =m_{H_0}^2-m_{A_0}^2 $. Note that the summation of all these diagrams effectively make the integral in Eq.~(\ref{TriangleSum_k}) finite. By multiplying various common factors (propagators and vertices) in the cocktail diagrams, we get
\begin{eqnarray}\label{total}
(-im_\nu)_{ab} &=& \frac{s_{2\beta}}{(16\pi^2)^3} (x_a C_{ab} x_b) \Big(\kappa_2 s_{2\beta} \frac{ (\Delta m_+^2)^2 \Delta m_0^2}{m_\rho^4 v^2} \Big) \nonumber\\
&& \frac{1}{2}(16\pi^2)^3 m_\rho^4 m_W^4 \gamma_\beta \gamma_\alpha \int \frac{d^4 k_2}{(2\pi)^4} \int \frac{d^4 k_1}{(2\pi)^4} \Big\{(g^{\alpha\mu}- \frac{k_{1}^{\alpha}k_{1}^{\mu}}{m_W^2}) (g^{\beta\nu}-\frac{k_{2}^{\beta}k_{2}^{\nu}}{m_W^2})\Big\}/\nonumber\\
&&\Big\{(k_2^2-m_W^2)(k_1^2-m_W^2) (k_2^2-m_a^2)(k_1^2-m_b^2)[(k_1+k_2)^2-m_\rho^2]\Big\}\nonumber\\
&& \int\frac{d^4 k}{(2\pi)^4} \Big\{(2k-k_2)_\nu (-2k-k_1)_\mu \Big\}/\Big\{(k^2-m_{H_0}^2)(k^2-m_{A_0}^2)[(k+k_1)^2-m_{H_1}^2]\nonumber\\
&& [(k+k_1)^2-m_{H_2}^2][(k-k_2)^2-m_{H_1}^2][(k-k_2)^2-m_{H_2}^2]\Big\}\, \nonumber\\
&=& (x_a C_{ab} x_b) \frac{s_{2\beta}}{(16\pi^2)^3} {\cal A}_{1\kappa_2} {\cal I}_1\, ,
\end{eqnarray}
where we have used $x_a= m_a/v$ and $m_W= g_2 v/\sqrt{2}$ and rearranged the factors for convenience.
Note that the prefactor ${\cal A}_{1\kappa_2}$ in the big parenthesis is precisely the first part of ${\cal A}_1$ in Eq.~(\ref{Mass}) and we denote ${\cal I}_{1}$ for the three-loop integral shown in the last four lines of the first equality.

\subsubsection{Integration Over $k$}
We now integrate the internal momentum $k$ in Eq.~(\ref{total}). By using the Feynman parameters $s_i$
to combine the three pairs of the propagators with the same momentum, we have
\begin{eqnarray}\label{CombineS}
{\rm I}_1 &=& \int\frac{d^4 k}{(2\pi)^4} \Big\{(2k-k_2)_\nu (-2k-k_1)_\mu\Big\}/\Big\{(k^2-m_{H_0}^2)(k^2-m_{A_0}^2)\nonumber\\
&& [(k+k_1)^2-m_{H_1}^2][(k+k_1)^2-m_{H_2}^2][(k-k_2)^2-m_{H_1}^2][(k-k_2)^2-m_{H_2}^2]\Big\} \nonumber\\
&=& \int^1_0 \prod_{i=1}^3 d s_i \int\frac{d^4 k}{(2\pi)^4} \frac{(2k-k_2)_\nu (-2k-k_1)_\mu}{(k^2-m_{s_3}^2)^2[(k+k_1)^2-m_{s_1}^2]^2[(k-k_2)^2-m_{s_2}^2]^2},
\end{eqnarray}
where we have defined $m_{s_1}^2=(1-s_1)m_{H_1}^2+s_1 m_{H_2}^2$,
$m_{s_2}^2=(1-s_2)m_{H_1}^2+s_2 m_{H_2}^2$ and
$m_{s_3}^2=(1-s_3)m_{H_0}^2+s_3 m_{A_0}^2$.
The combination of the remaining three factors in the denominator above gives
\begin{eqnarray}
{\rm I}_1 &=& \int^1_0 \prod_{i=1}^3 d s_i \Gamma(6)\int d x_1 d x_2 \int \frac{d^4 k}{(2\pi)^4} \{ x_1 x_2 (1-x_1-x_2)\,(2k-k_2)_\nu (-2k-k_1)_\mu \}/\nonumber\\
&& {\Big(x_1[(k+k_1)^2-m_{s_1}^2] +x_2[(k-k_2)^2-m_{s_2}^2]+(1-x_1-x_2)[k^2-m_{s_3}^2] \Big)^6}\nonumber\\
&=& \int^1_0 \prod_{i=1}^3 d s_i \Gamma(6)\int d x_1 d x_2 x_1 x_2 (1-x_1-x_2)\nonumber\\
&& \int \frac{d^4 k}{(2\pi)^4}
\frac{-4k_\mu k_\nu +[-(1-2 x_1)k_1-2x_2k_2]_\mu[-2x_1 k_1+(2x_2-1)k_2]_\nu}{(k^2-m_x^2)^6}\;,
\end{eqnarray}
where we have made the translation of the momentum $k\rightarrow k-x_1k_1+x_2k_2$ and defined
\begin{equation}
m^2_x=\sum_i^{i=3} x_i m_{s_i}^2-x_1(1-x_1)k_1^2 - 2 x_1 x_2 k_1\cdot k_2 - x_2(1-x_2) k_2^2\, ,
\end{equation}
with $x_3=1-x_1-x_2$. We have also ignored the terms proportional to the odd powers of $k$ since they vanish after the integration. The integration over $k$ leads to
\begin{eqnarray}
{\rm I}_1 &=& \int^1_0 \prod_{i=1}^3 d s_i \int d x_1 d x_2 x_1 x_2(1-x_1-x_2) \frac{i}{16\pi^2}\Gamma(3)\Big( \frac{2 g_{\mu\nu}}{(m_x^2)^3}
 + \frac{3 N_{\mu\nu}}{(m_x^2)^4} \Big),\nonumber\\
 &=& {\rm I}_{11} +{\rm I}_{12}\, ,
\end{eqnarray}
where
\begin{equation}\label{DefN}
N_{\mu\nu}=[-(1-2 x_1)k_1-2x_2k_2]_\mu[-2x_1 k_1+(2x_2-1)k_2]_\nu\, .
\end{equation}

\subsubsection{Integration of ${\rm I}_{11}$ Over $k_1$ and $k_2$ }
The integration over $k_1$ for ${\rm I}_{11}$ is defined as
\begin{eqnarray}
{\rm II}_{1} = \frac{i}{16\pi^2} \Gamma(3) x_1 x_2 (1-x_1-x_2) \int \frac{d^4 k_1}{(2\pi)^4} \frac{[g^{\alpha\mu }-k_1^\alpha k_1^\mu /m_W^2]}{(k_1^2-m_W^2)(k_1^2-m_b^2)[(k_1+k_2)^2-m_\rho^2]} \frac{2 g_{\mu\nu}}{(m_x^2)^3}\, ,\nonumber\\
\label{EQintk1}
\end{eqnarray}
where we have suppressed the integration measure for the Feynman parameters $x_j$ and $s_i$ to simplify our formulae.
With the Feynman parameters $z_i$ ($i=1,...3$), we can combine all the factors in the denominator
\begin{eqnarray}\label{II1}
{\rm II}_{1} = \frac{i\Gamma(3)}{16\pi^2} \frac{ (-1)x_1 x_2 (1-x_1-x_2)}{x_1^3 (1-x_1)^3} \frac{\Gamma(6)}{\Gamma(3)}\int \prod^3_{j=3} d z_j \frac{2z_1^2[g^\alpha_\nu-k_1^\alpha k_{1\nu} /m_W^2]}{D^6} \, ,
\end{eqnarray}
where
\begin{eqnarray}\label{D_exp}
D &=& z_1\Big[k_1^2+\frac{2x_1 x_2}{x_1(1-x_1)}k_1\cdot k_2 + \frac{x_2(1-x_2)}{x_1(1-x_1)}k_2^2 - \frac{\sum_i x_i m_{s_i}^2}{x_1(1-x_1)}\Big] +z_2 \Big[(k_1+k_2)^2 -m_\rho^2\Big]
\nonumber\\
&& +z_3(k_1^2-m_W^2) +(1-z_1-z_2-z_3)(k_1^2-m_b^2)
\end{eqnarray}

With the internal momentum translation: $k_1\rightarrow k_1- c_2 k_2$ where $c_2 = [x_2 z_1+(1-x_1)z_2]/(1-x_1)$ and
the integration of $k_1$, we obtain
\begin{eqnarray}\label{II1p}
{\rm II}_{1} &=& \frac{1}{(16\pi^2)^2} \int \prod_{j=1}^3 d z_j \frac{x_1 x_2 (1-x_1-x_2)z_1^2}{x_1^3 (1-x_1)^3} \nonumber\\
&&  \Big\{ \frac{2\Gamma(4) (g^\alpha_\nu - c_2^2 k_{2}^\alpha k_{2\nu}/m_W^2)}{B_2^4 [k_2^2 - \Delta]^4} - \frac{\Gamma(3)}{m_W^2} \frac{g^\alpha_\nu}{B_2^3 [k_2^2 - \Delta]^3} \Big\}\, ,
\end{eqnarray}
where
\begin{eqnarray}
B_2 &=& \frac{x_2(1-x_2)z_1}{x_1(1-x_1)} + z_2 -c_2^2 \, ,\label{B2Func}\\
\Delta &=& \frac{z_2}{B_2}m_\rho^2 + \frac{z_3}{B_2}m_W^2 + \frac{1-z_1-z_2-z_3}{B_2} m_b^2 + \frac{z_1\sum_i x_i m_{s_i}^2}{B_2x_1(1-x_1)}\, .
\end{eqnarray}
The integration over $k_2$ can be similarly done and the result is given by
\begin{eqnarray}
{\rm III}_{1} &=& \frac{\Gamma(3)}{(16\pi^2)^2} \frac{x_1 x_2 (1-x_1-x_2) z_1^2}{x_1^3 (1-x_1)^3} \int \frac{d^4 k_2}{(2\pi)^4} \frac{(g^{\beta\nu}-k_2^\beta k_2^\nu / m_W^2)}{(k_2^2-m_W^2)(k_2^2-m_a^2)}\nonumber\\
&& \Big\{ \frac{6 (g^\alpha_\nu - c_2^2 k_{2}^\alpha k_{2\nu}/m_W^2)}{B_2^4 [k_2^2 - \Delta]^4} - \frac{1}{m_W^2} \frac{g^\alpha_\nu}{B_2^3 [k_2^2 - \Delta]^3} \Big\}\, ,
\end{eqnarray}
where the integration measures for the Feynman parameters are also suppressed for simplicity.
By combining the factors in the denominator with the Feynman parameters $y_i$, the expression can be transformed into
\begin{eqnarray}\label{y_para}
{\rm III}_{1} &=& \frac{\Gamma(3)}{(16\pi^2)^2} \frac{x_1 x_2 (1-x_1-x_2)z_1^2}{x_1^3 (1-x_1)^3} \int d y_1 d y_2 \int \frac{d^4 k_2}{(2\pi)^4} (g^{\beta\nu}- k_2^\beta k_2^\nu/m_W^2)\nonumber\\
&& \Big\{ \frac{\Gamma(6)}{\Gamma(4)}\frac{6y_1^3}{B_2^4} \frac{(g_\nu^\alpha - c_2^2 k_2^\alpha k_{2\nu}/m_W^2)}{(k_2^2 - [m_\rho^2 \Sigma]/[B_2 x_1 (1-x_1)])^6} - \frac{\Gamma(5)}{\Gamma(3)}\frac{y_1^2}{m_W^2 B_2^3} \frac{g^\alpha_\nu}{(k_2^2 - [m_\rho^2\Sigma]/[B_2 x_1 (1-x_1)])^5} \Big\}\, , \nonumber\\
\end{eqnarray}
where
\begin{eqnarray}
\Sigma &=& x_1(1-x_1)\Big[y_1 z_2 + (y_1 z_3 + y_2 B_2)\frac{m_W^2}{m_\rho^2} + y_1 (1-z_1-z_2-z_3)\frac{m_b^2}{m_\rho^2} +(1-y_1-y_2) B_2 \frac{m_a^2}{m_\rho^2}\Big] \nonumber\\
&& + y_1 z_1 \frac{(\sum_i x_i m_{s_i}^2)}{m_\rho^2} \,.\nonumber\\
\end{eqnarray}
After integrating out $k_2$, we find
\begin{eqnarray}\label{I31Final}
{\rm III}_{1} &=& \frac{i}{(16\pi^2)^3}{[x_1 x_2 (1-x_1-x_2)z_1^2]} \frac{g^{\alpha\beta}}{4} \Big\{ y_1^3\Big[\frac{48 x_1 (1-x_1)}{m_\rho^8\Sigma^4} + \frac{8(c_2^2+1)}{m_W^2 m_\rho^6 B_2 \Sigma^3} \nonumber\\
&& + \frac{12 c_2^2}{m_W^4 m_\rho^4 B_2^2 x_1(1-x_1) \Sigma^2 }\Big] +y_1^2 \Big[\frac{8}{m_W^2 m_\rho^6 \Sigma^3} + \frac{2}{m_W^4 m_\rho^4 B_2 x_1 (1-x_1) \Sigma^2}\Big] \Big\}\, .
\end{eqnarray}

\subsubsection{Integration of ${\rm I}_{12}$ Over $k_1$ and $k_2$}
The integration over $k_1$ for ${{\rm I}_{12}}$ can be written as
\begin{eqnarray}
{\rm II}_2 = \frac{i}{16\pi^2} \Gamma(3) x_1 x_2 (1-x_1-x_2) \int\frac{d^4 k_1}{(2\pi)^4}
\frac{(g^{\alpha\mu}-k_1^\alpha k_1^\mu / m_W^2)}{(k_1^2-m_W^2)(k_1^2-m_b^2)[(k_1+k_2)^2-m_\rho^2]} \frac{3 N_{\mu\nu}}{(m^2_x)^4} \, , \nonumber\\
\end{eqnarray}
where we have also suppressed the Feynman parameter integration measures. 
Similar to the derivation of Eq.~(\ref{II1}) from 
Eq.~(\ref{EQintk1}), we have
\begin{eqnarray}
{\rm II}_2 = \frac{i}{16\pi^2} \frac{x_1 x_2 (1-x_1-x_2)}{x_1^4 (1-x_1)^4}{\Gamma(7)} z_1^3 \int\frac{d^4 k_1}{(2\pi)^4} \frac{N^{\prime \alpha}_\nu}{D^7}\, ,
\end{eqnarray}
where $D$ is defined in Eq.~(\ref{D_exp}) and
\begin{eqnarray}\label{Def_N}
N'^\alpha_\nu &=&
[-(1-2x_1)(k_1-c_2k_2)-2x_2k_2]_\mu[-2x_1(k_1-c_2k_2)+(2x_2-1)k_2]_\nu \nonumber\\
&& [g^{\alpha\mu}-{(k_1-ck_2)^\alpha (k_1-c_2k_2)^\mu}/{m_W^2}]\nonumber\\
&=& (1-2x_1)(2x_1)k_{1}^\alpha k_{1\nu}+d_1d_2k_{2}^\alpha k_{2\nu} - (1-2x_1)(2x_1)\frac{1}{m_W^2}k_1^2k_{1}^\alpha k_{1\nu} \nonumber\\
&& -(1-2x_1)(2x_1)c_2^2\frac{1}{m_W^2}k_1\cdot k_2 k_{2}^\alpha k_{1\nu}
 -(1-2x_1)d_2c_2 \frac{1}{m_W^2}k_1\cdot k_2 k_1^\alpha k_{2\nu} \nonumber\\
&& -(1-2x_1)d_2c_2\frac{1}{m_W^2}k_1^2k_2^\alpha k_{2\nu}
 -(2x_1)d_1c_2\frac{1}{m_W^2}k_2^2 k_1^\alpha k_{1\nu} \nonumber\\
&& -(2x_1)d_1c_2\frac{1}{m_W^2}k_1\cdot k_2 k_2^\alpha k_{1\nu} -d_1d_2\frac{1}{m_W^2}k_1\cdot k_2 k_1^\alpha k_{2\nu}-d_1d_2c_2^2\frac{1}{m_W^2}k_2^2 k_{2}^\alpha k_{2\nu}\;,
\end{eqnarray}
with
\begin{eqnarray}\label{Def_dd}
d_1=c_2-2c_2\,x_1-2\,x_2\;,d_2=2\,x_1c_2+2\,x_2-1\;.
\end{eqnarray}
The integration over $k_1$ can be easily carried out with the result given by
\begin{eqnarray}
{\rm II}_2&=&\frac{\Gamma(3)}{(16\pi^2)^2}\frac{x_1x_2(1-x_1x_2)z_1^3}{x_1^4(1-x_1)^4}\Big\{
-\frac{12d_1d_2 k_2^\alpha k_{2\nu}}{B_2^5[k_2^2-\Delta]^5}
+\frac{12d_1d_2c_2^2 k_2^2 k_2^\alpha k_{2\nu}}{B_2^5m_W^2[k_2^2-\Delta]^5}
-\frac{6(1-2\,x_1)(2\,x_1)g_{\nu}^\alpha}{ 4B_2^4[k_2^2-\Delta]^4}\nonumber\\
&&+\frac{6(1-2\,x_1)(2\,x_1)c_2^2k_{2\nu}k_{2}^\alpha}{4B_2^4m_W^2[k_2^2-\Delta]^4}
+\frac{6(1-2\,x_1)d_2c_2k_{2\nu}k_{2}^\alpha}{4B_2^4m_W^2[k_2^2-\Delta]^4}
+\frac{6(1-2\,x_1)d_2c_2 k_{2\nu}k_{2}^\alpha}{B_2^4 m_W^2[k_2^2-\Delta]^4}\nonumber\\
&&+\frac{6(2\,x_1)d_1c_2g_{\nu}^\alpha k_{2}^2}{4B_2^4m_W^2[k_2^2-\Delta]^4}
+\frac{6(2\,x_1)d_1c_2 k_{2\nu}k_2^\alpha}{4B_2^4m_W^2[k_2^2-\Delta]^4}
+\frac{6d_1d_2k_{2\nu}k_2^\alpha}{4B_2^4m_W^2[k_2^2-\Delta]^4}
+\frac{6(1-2\,x_1)(2\,x_1)g_{\nu}^\alpha}{4B_2^3m_W^2[k_2^2-\Delta]^3}
\Big\}\;. \nonumber\\
\end{eqnarray}
By appending the rest propagators involving $k_2$ and performing the Feynman parametrization with $y_i$ as that in Eq.~(\ref{y_para}), the expression becomes
\begin{eqnarray}
{\rm III_2}&=& \int\frac{d^4k_2}{(2\pi)^4}
{({g^{\beta\nu}-\frac{k_2^\beta k_2^\nu}{m_W^2}})\over (k_2^2-m_W^2)(k_2^2-m_a^2)}{\rm II}_2\nonumber\\
&=&{1\over(16\pi^2)^2}{x_1x_2(1-x_1-x_2)z_1^3\over x_1^4(1-x_1)^4}\int{d^4k_2\over(2\pi)^4}\Big\{
-{\Gamma(7)d_1d_2 y_1^4\over B_2^5}{{k_2^\alpha k_2^\beta(1-{k_2^2\over m_W^2}})\over [k_2^2-(m_\rho^2\Sigma)/ (x_1(1-x_1)B_2)]^7}\nonumber\\
&&+{\Gamma(7)d_1d_2c_2^2 y_1^4\over B_2^5}{k_2^\alpha k_2^\beta k_2^2(1-{k_2^2\over m_W^2})
\over m_W^2[k_2^2-(m_\rho^2\Sigma)/ (x_1(1-x_1)B_2)]^7}\nonumber\\
&&
-{\Gamma(6)(1-2x_1)(2x_1)y_1^3\over 2B_2^4}{(g^{\alpha\beta}-{k_{2}^\alpha k_2^\beta\over m_W^2})
\over [k_2^2-(m_\rho^2\Sigma)/ (x_1(1-x_1)B_2)]^6}\nonumber\\
&&+{\Gamma(6)(1-2x_1)(2x_1)c_2^2y_1^3\over 2B_2^4}{k_2^\alpha k_2^\beta (1-{k_2^2\over m_W^2})
\over m_W^2[k_2^2-(m_\rho^2\Sigma)/ (x_1(1-x_1)B_2)]^6}\nonumber\\
&&+{\Gamma(6)(1-2x_1)d_2c_2y_1^3\over 2B_2^4}{k_2^\alpha k_2^\beta (1-{k_2^2\over m_W^2})
\over m_W^2[k_2^2-(m_\rho^2\Sigma)/ (x_1(1-x_1)B_2)]^6}\nonumber\\
&&+{2\Gamma(6)(1-2x_1)d_2c_2y_1^3\over B_2^4}{k_2^\alpha k_2^\beta (1-{k_2^2\over m_W^2})
\over m_W^2[k_2^2-(m_\rho^2\Sigma)/ (x_1(1-x_1)B_2)]^6}\nonumber\\
&&+{\Gamma(6)(2x_1)d_1c_2y_1^3\over 2B_2^4}{ k_2^2(g^{\alpha\beta}-{k_2^\alpha k_2^\beta\over m_W^2})
\over m_W^2[k_2^2-(m_\rho^2\Sigma)/ (x_1(1-x_1)B_2)]^6}\nonumber\\
&&+{\Gamma(6)(2x_1)d_1c_2y_1^3\over 2B_2^4}{ k_2^\alpha k_2^\beta(g^{\alpha\beta}-{k_2^2\over m_W^2})
\over m_W^2[k_2^2-(m_\rho^2\Sigma)/ (x_1(1-x_1)B_2)]^6}\nonumber\\
&&
+{\Gamma(6)d_1d_2y_1^3\over 2B_2^4}{ k_2^\alpha k_2^\beta(1-{k_2^2\over m_W^2})
\over m_W^2[k_2^2-(m_\rho^2\Sigma)/ (x_1(1-x_1)B_2)]^6}\nonumber\\
&&+{6\Gamma(5)(1-2x_1)(2x_1)y_1^2\over 4B_2^3}{ g^{\alpha\beta}-{k_2^\alpha k_2^\beta\over m_W^2}
\over m_W^2[k_2^2-(m_\rho^2\Sigma)/ (x_1(1-x_1)B_2)]^5}
\Big\}\;.\nonumber\\
\end{eqnarray}

Finally, the integration over $k_2$ gives
\begin{eqnarray}
{\rm III_2}&=&{i\over (16\pi^2)^3}x_1x_2(1-x_1-x_2)z_1^3{g^{\alpha\beta}\over 4}\{
-d_1d_2y_1^4({12\over B_2\Sigma^4m_\rho^8}+{12\over x_1(1-x_1)B_2^2\Sigma^3m_\rho^6 m_W^2})\nonumber\\
&& -d_1d_2c_2^2y_1^4({12\over x_1(1-x_1)B_2^2\Sigma^3 m_\rho^6 m_W^2}+\frac{24}{x_1^2(1-x_1)^2B_2^3\Sigma^2 m_\rho^4 m_W^4}) \nonumber\\
&& -2x_1(1-2x_1)y_1^3({12\over \Sigma^4m_\rho^8}+{2\over x_1(1-x_1)B_2\Sigma^3m_\rho^6m_W^2})\nonumber\\
&& -2x_1(1-2x_1)c_2^2y_1^3({2\over x_1(1-x_1)B_2\Sigma^3m_\rho^6m_W^2}+{3\over x_1^2(1-x_1)^2B_2^2\Sigma^2m_\rho^4m_W^4})\nonumber\\
&& -(1-2x_1)d_2c_2y_1^3({2\over x_1(1-x_1)B_2\Sigma^3m_\rho^6m_W^2}+{3\over x_1^2(1-x_1)^2B_2^2\Sigma^2m_\rho^4m_W^4})\nonumber\\
&& -(1-2x_1)d_2c_2y_1^3({8\over x_1(1-x_1)B_2\Sigma^3m_\rho^6m_W^2}+{12\over x_1^2(1-x_1)^2B_2^2\Sigma^2m_\rho^4m_W^4})\nonumber\\
&& -2x_1d_1c_2y_1^3({8\over x_1(1-x_1)B_2\Sigma^3m_\rho^6m_W^2}+{3\over x_1^2(1-x_1)^2B_2^2\Sigma^2m_\rho^4m_W^4})\nonumber\\
&& -2x_1d_1c_2y_1^3({2\over x_1(1-x_1)B_2\Sigma^3m_\rho^6m_W^2}+{3\over x_1^2(1-x_1)^2B_2^2\Sigma^2m_\rho^4m_W^4})\nonumber\\
&& -d_1d_2y_1^3({2\over x_1(1-x_1)B_2\Sigma^3m_\rho^6m_W^2}+{3\over x_1^2(1-x_1)^2B_2^2\Sigma^2m_\rho^4m_W^4})\nonumber\\
&& -(1-2x_1)(2x_1)y_1^2({12\over x_1(1-x_1)\Sigma^3m_\rho^6m_W^2}+{3\over x_1^2(1-x_1)^2B_2\Sigma^2m_\rho^4m_W^4})
\}\;.
\end{eqnarray}

\subsubsection{Final Results}
By summing up the above two integration results and multiplying the prefactors, we obtain
\begin{eqnarray}\label{result}
{\cal I}_1 = \frac{1}{2}(16\pi^2)^3 m_\rho^4 m_W^4 \gamma_\beta \gamma_\alpha ({\rm III}_1 + {\rm III}_2)
= \frac{i}{2} (\frac{m_W^4}{m_\rho^4}{\cal I}_{12} + \frac{m_W^2}{m_\rho^2}{\cal I}_{11} + {\cal I}_{10} )\,,
\end{eqnarray}
with
\begin{eqnarray}
{\cal I}_{10} &=& \frac{x_1 x_2 (1-x_1-x_2) z_1^2}{\Sigma^4} \Big(48 x_1 (1-x_1)y_1^3 - \frac{12 d_1 d_2\, y_1^4 z_1}{B_2} -12(1-2 x_1)(2x_1)y_1^3 z_1 \Big)\, ,\,\,\,\,\,\,
\end{eqnarray}
\begin{eqnarray}
{\cal I}_{11} &=& \frac{x_1 x_2 (1-x_1-x_2) z_1^2}{\Sigma^3} \Big(\frac{8 y_1^3 (c_2^2+1)}{B_2} + 8 y_1^2\Big) + \frac{x_1 x_2 (1-x_1-x_2) z_1^3}{x_1 (1-x_1 )\Sigma^3}\Big( -\frac{12 d_1 d_2\, y_1^4}{B_2^2}\nonumber\\
&&  -\frac{12 d_1 d_2 c_2^2 y_1^4}{B_2^2} - \frac{2(1-2x_1)(2x_1)y_1^3}{B_2} - \frac{2(1-2 x_1)(2 x_1) c_2^2 y_1^3}{B_2} - \frac{2(1-2 x_1)d_2 c_2 y_1^3}{B_2}  \nonumber\\
&& - \frac{8 (1-2 x_1) d_2 c_2 y_1^3}{B_2} - \frac{8 (2 x_1) d_1 c_2 y_1^3 }{B_2} - \frac{2 (2 x_1) d_1 c_2 y_1^3}{B_2} -\frac{2 d_1 d_2 y_1^3}{B_2} -12 (1-2 x_1)(2x_1)y_1^2 \Big) \, ,\nonumber\\
\end{eqnarray}
\begin{eqnarray}
{\cal I}_{12} &=& \frac{x_1 x_2 (1-x_1- x_2)z_1^2}{x_1 (1-x_1) \Sigma^2} \Big( \frac{12 y_1^3 c_2^2}{B_2^2} + \frac{2 y_1^2}{B_2} \Big) + \frac{x_1 x_2 (1-x_1-x_2) z_1^3}{x_1^2 (1-x_1)^2 \Sigma^2} \Big( -\frac{24 d_1 d_2 c_2^2 y_1^4}{B_2^3} \nonumber\\
&&  -\frac{3(1-2 x_1)(2 x_1) c_2^2 y_1^3}{B_2^2} - \frac{3 (1-2 x_1) d_2 c_2 y_1^3}{B_2^2} -\frac{12 (1-2 x_1) d_2 c_2 y_1^3}{B_2^2} -\frac{3 (2 x_1) d_1 c_2 y_1^3}{B_2^2} \nonumber\\
&& -\frac{3 (2 x_1) d_1 c_2 y_1^3}{B_2^2} - \frac{3 d_1 d_2 y_1^3}{B_2^2} -\frac{3 (1-2x_1)(2x_1) y_1^2}{B_2} \Big) ,
\end{eqnarray}
where the final results are classified according to the powers of $m_W^2/m_\rho^2$.
Here,
we have suppressed the integration measures for the Feynman parameters $x_i,\, y_i,\, z_i$ and $s_i$, defined by
\begin{eqnarray}
{\rm measure} &=& \int^1_0 ds_1 \int^1_0 ds_2 \int^1_0 ds_3 \int^1_0 d x_1 \int^{1-x_1}_0 dx_2 \int^1_0 dy_1 \int^{1-y_1}_0 dy_2 \nonumber\\
&& \int^1_0 dz_1 \int^{1-z_1}_0 dz_2 \int^{1-z_1-z_2}_0 dz_3\,.
\end{eqnarray}
It is clear that this complicated 10-dimensional Feynman parameter integration can only be calculated with the help of a numerical package.
In our work, we use three widely-applied numerical integration softwares: {\sf Mathematica}(Global Adaptive), {\sf SecDec-2.1.4}~\cite{SecDec} and VEGAS in {\sf GSL}~\cite{GSL} in order to cross-check the accuracy and stability of the calculation. With Eq.~(\ref{result}), we can study the benchmark point: 
\begin{eqnarray}
m_{H^0}&=&70~{\rm GeV}\,, \;m_{A^0}=250~{\rm GeV}\,,\; m_{H_1^+}=90~{\rm GeV}\,,\; m_{H_2^+}=400~{\rm GeV}\,,\; m_{\rho}=1~{\rm TeV}\,,\nonumber\\ C_{e\tau}&=&0.06\,,\; C_{\mu\mu}=0.01\,,\; C_{\mu\tau}=0.0009\,,\; C_{\tau\tau}=5\times 10^{-5}\,,\kappa_2=2~{\rm TeV}
\label{Eq_BMpoint}
\end{eqnarray}
in the first version of Ref.~\cite{Gustafsson:2012vj} before its Erratum with $\xi=0$, and the final results are given by
\begin{eqnarray}
{\rm Mathematica:} &\quad & {\cal I}_{10} = 7264.5\pm 104.4 , \quad {\cal I}_{11} = 124.667\pm 1.818 , \quad {\cal I}_{12} = 4.10278 \pm 0.0234 ;\,\nonumber\\
&& {\cal I}_1 = 2.61 i, \quad m_{\nu} =
\left(
\begin{matrix}
{\mathcal O}(10^{-3}) & {\mathcal O}(10^{-3}) & 1.05 \\
{\mathcal O}(10^{-3}) & 2.16 & 3.25 \\
1.05 & 3.25 & 3.03
 \end{matrix}
\right)\times 10^{-13}~~\mathrm{GeV}, \\
{\rm GSL-VEGAS:} &\quad & {\cal I}_{10} = 7350.752 \pm 5.271,\quad {\cal I}_{11} = 125.122 \pm 0.116,\quad {\cal I}_{12} = 4.107\pm 0.003; \,\nonumber\\
&& {\cal I}_1 = 2.612 i, \quad m_{\nu} =
\left(
\begin{matrix}
{\mathcal O}(10^{-3}) & {\mathcal O}(10^{-3}) & 1.05 \\
{\mathcal O}(10^{-3}) & 2.16 & 3.26 \\
1.05 & 3.26 & 3.04
 \end{matrix}
\right)\times 10^{-13}~~\mathrm{GeV},\\
\textrm{ SecDec-2.1.4 :} &\quad & {\cal I}_{10} = 7353.2 \pm 7.3,\quad {\cal I}_{11} = 125.79 \pm 0.04,\quad {\cal I}_{12} = 4.108\pm 0.001; \,\nonumber\\
&& {\cal I}_1 = 2.612 i, \quad m_{\nu} =
\left(
\begin{matrix}
{\mathcal O}(10^{-3}) & {\mathcal O}(10^{-3}) & 1.05 \\
{\mathcal O}(10^{-3}) & 2.16 & 3.26 \\
1.05 & 3.26 & 3.04
 \end{matrix}
\right)\times 10^{-13}~~\mathrm{GeV}.
\end{eqnarray}
As expected, the numerical values from the three packages are essentially the same. Clearly, the obtained neutrino masses with $\xi=0$ are smaller than the experimental values by about two orders. 



\subsection{Integrals proportional to $\xi v$}
In this subsection, we go on to calculate the neutrino mass part which is proportional to $\xi v$.The relevant pieces of the Lagrangian are
\begin{eqnarray}\label{Lag_xi}
\mathcal{L}_\xi &=& - \xi \Phi_2^T i\sigma_2 \Phi_1 S^+ \rho^{--} +{\rm h.c.} = - \xi v \Lambda^+ S^+ \rho^{--} +{\rm h.c.}\nonumber\\
& = & -\xi v \rho^{--}[s_\beta c_\beta H_1^+ H_1^+ + (c_\beta^2-s_\beta^2) H_1^+H_2^+ - s_\beta c_\beta H_2^+H_2^+]+\mathrm{h.c.}.
\end{eqnarray}
These vertices, together with the first term in Eq.~(\ref{Lag_k}), also give 8 Feynman diagrams in the unitary gauge with  the upper triangle loops as their only differences. Among them, the four triangle factors with $H_0$ running inside are
\begin{eqnarray}
(H_0H_1H_1)&:& (\xi v)\frac{g_2^2 s_{2\beta} (c_{2\beta}+1)}{8} \frac{(2k-k_2)_\nu (-2k-k_1)_\mu}{(k^2-m_{H_0}^2)[(k+k_1)^2-m_{H_1}^2][(k-k_2)^2-m_{H_1}^2]}\;,\nonumber\\
(H_0H_1H_2)&:&-(\xi v)\frac{g_2^2 s_{2\beta}c_{2\beta}}{8} \frac{(2k-k_2)_\nu (-2k-k_1)_\mu}{(k^2-m_{H_0}^2)[(k+k_1)^2-m_{H_1}^2][(k-k_2)^2-m_{H_2}^2]}\;,\nonumber\\
(H_0H_2H_1)&:&-(\xi v)\frac{g_2^2 s_{2\beta}c_{2\beta}}{8} \frac{(2k-k_2)_\nu (-2k-k_1)_\mu}{(k^2-m_{H_0}^2)[(k+k_1)^2-m_{H_2}^2][(k-k_2)^2-m_{H_1}^2]}\;,\nonumber\\
(H_0H_2H_2)&:&(\xi v)\frac{g_2^2 s_{2\beta} (c_{2\beta}-1)}{8} \frac{(2k-k_2)_\nu (-2k-k_1)_\mu}{(k^2-m_{H_0}^2)[(k+k_1)^2-m_{H_2}^2][(k-k_2)^2-m_{H_2}^2]}\;,
\end{eqnarray}
while an extra overall minus sign should be multiplied for the corresponding formulae involving the pseudoscalar $A_0$. Thus, the summation of all these 8 diagrams yields
\begin{eqnarray}
&& (\xi v)\frac{g_2^2 s_{2\beta} c_{2\beta}}{8}\int\frac{d^4 k}{(2\pi)^4}\Big\{(2k-k_2)_\nu (-2k-k_1)_\mu (\Delta m_+^2)^2 \Delta m_0^2 \Big\}/\Big\{(k^2-m_{H_0}^2)(k^2-m_{A_0}^2)\nonumber\\
&& [(k+k_1)^2-m_{H_1}^2][(k+k_1)^2-m_{H_2}^2][(k-k_2)^2-m_{H_1}^2][(k-k_2)^2-m_{H_2}^2]\Big\} \nonumber \\
&& + (\xi v) \frac{g_2^2 s_{2\beta}}{8} \int\frac{d^4 k}{(2\pi)^4} \nonumber\\
&& \Big\{ \frac{(2k-k_2)_\nu (-2k-k_1)_\mu \Delta m_+^2 \Delta m_0^2}{(k^2-m_{H_0}^2)(k^2-m_{A_0}^2)[(k+k_1)^2-m_{H_1}^2][(k-k_2)^2-m_{H_1}^2][(k-k_2)^2-m_{H_2}^2]} \nonumber\\
&&  +\frac{(2k-k_2)_\nu (-2k-k_1)_\mu \Delta m_+^2 \Delta m_0^2}{(k^2-m_{H_0}^2)(k^2-m_{A_0}^2)[(k+k_1)^2-m_{H_1}^2][(k+k_1)^2-m_{H_2}^2][(k-k_2)^2-m_{H_2}^2]} \Big\} \,.
\end{eqnarray}
Note that the integral in the first two lines is the same as that proportional to $\kappa_2$ in Eq.~(\ref{TriangleSum_k}), so we expect that it gives the same result ${\cal I}_1$. Thus, the contribution to neutrino masses proportional to $\xi$ can be written as
\begin{eqnarray}\label{total_xi}
(-im_\nu)_{ab} = (x_a C_{ab} x_b) \frac{s_{2\beta}}{(16\pi^2)^3} ({\cal A}_{1\xi} {\cal I}_1 + {\cal A}_2 {\cal I}_2 )
\end{eqnarray}
where,
\begin{eqnarray}
{\cal A}_{1\xi} = \xi v c_{2\beta} \frac{ (\Delta m_+^2)^2 \Delta m_0^2}{m_\rho^4 v^2}, \quad\quad
{\cal A}_2 = \frac{\xi v}{m_\rho^2} \frac{\Delta m_+^2 \Delta m_0^2}{v^2}
\end{eqnarray}
\begin{eqnarray}\label{IntTotal_xi}
{\cal I}_2 &=& \frac{1}{2}(16\pi^2)^3 m_\rho^2 m_W^4 \gamma_\beta \gamma_\alpha \int \frac{d^4 k_2}{(2\pi)^4} \int \frac{d^4 k_1}{(2\pi)^4} \nonumber\\
&& \frac{(g^{\alpha\mu}- \frac{k_{1}^{\alpha}k_{1}^{\mu}}{m_W^2}) (g^{\beta\nu}-\frac{k_{2}^{\beta}k_{2}^{\nu}}{m_W^2})}{(k_2^2-m_W^2)(k_1^2-m_W^2)(k_2^2-m_a^2)(k_1^2-m_b^2)[(k_1+k_2)^2-m_\rho^2]}\nonumber\\
&& \int\frac{d^4 k}{(2\pi)^4} \Big\{ \frac{(2k-k_2)_\nu (-2k-k_1)_\mu}{(k^2-m_{H_0}^2)(k^2-m_{A_0}^2)[(k+k_1)^2-m_{H_1}^2][(k-k_2)^2-m_{H_1}^2][(k-k_2)^2-m_{H_2}^2]} \nonumber\\
&&+\frac{(2k-k_2)_\nu (-2k-k_1)_\mu}{(k^2-m_{H_0}^2)(k^2-m_{A_0}^2)[(k+k_1)^2-m_{H_1}^2][(k+k_1)^2-m_{H_2}^2][(k-k_2)^2-m_{H_2}^2]} \Big\}\nonumber\\
&=& {\cal I}^\prime_{2} + {\cal I}^{\prime\prime}_2 .
\end{eqnarray}
Here, we have separated ${\cal I}_2$ into two parts, ${\cal I}_2^\prime$ and ${\cal I}_2^{\prime\prime}$, which are defined in the second and third lines in the first equality, respectively. Note that ${\cal I}_2^\prime$ and ${\cal I}_2^{\prime\prime}$ are symmetric to each other by the exchange of the charged scalar masses $m_{H_1}^2 \leftrightarrow m_{H_2}^2$. Thus, in practice, we only need to calculate ${\cal I}_2^\prime$, and find the result of ${\cal I}_2^{\prime\prime}$ with such a mass exchange, as is done in the following subsections.

\subsubsection{Integration Over $k$}
We first integrate $k$ in ${\cal I}^\prime_2$. With the Feynman parameters $s_i$ and $x_i$, we 
combine the propagators in the denominator as follows
\begin{eqnarray}
{\rm I}_1^\prime
&=& \int^1_0 \prod_{i=1}^3 d s_i \Gamma(5)\int d x_1 d x_2 x_2 (1-x_1-x_2)\nonumber\\
&& \int \frac{d^4 k}{(2\pi)^4}
\frac{-4k_\mu k_\nu +[-(1-2 x_1)k_1-2x_2k_2]_\mu[-2x_1 k_1+(2x_2-1)k_2]_\nu}{(k^2-m_x^{\prime 2})^5}\;,
\end{eqnarray}
where we have made the translation of the momentum $k\rightarrow k-x_1k_1+x_2k_2$ and defined
\begin{equation}
m^{\prime 2}_x= m_\sigma^2-x_1(1-x_1)k_1^2 - 2 x_1 x_2 k_1\cdot k_2 - x_2(1-x_2) k_2^2\, ,
\end{equation}
with $m_\sigma^2\equiv x_1 m_{H_1}^2+\sum_{i=2}^{3} x_i m_{s_i}^2 $ and $x_3=1-x_1-x_2$. The integration over $k$ can be subsequently performed with the following results
\begin{eqnarray}
{\rm I}^\prime_1 &=& \int^1_0 \prod_{i=2}^3 d s_i \int d x_1 d x_2 x_2(1-x_1-x_2) \frac{(-i)}{16\pi^2}\Big[ \frac{2 g_{\mu\nu}}{(m_x^{\prime 2})^2}
 + \frac{2 N_{\mu\nu}}{(m_x^{\prime 2})^3} \Big]\nonumber\\
 &=& {\rm I}^\prime_{11} +{\rm I}^\prime_{12}\, ,
\end{eqnarray}
where $N_{\mu\nu}$ is the same as in Eq.~(\ref{DefN}). We have separated ${\rm I}^\prime_1$ into ${\rm I}^\prime_{11}$ and ${\rm I}^\prime_{12}$ in terms of their powers of $m_x^{\prime 2}$.

\subsubsection{Integration of ${\rm I}^\prime_{11}$ Over $k_1$ and $k_2$}
The integration over $k_1$ for ${\rm I}^\prime_{11}$ is defined as
\begin{eqnarray}
{\rm II}^\prime_{1} = \frac{-i}{16\pi^2} x_2 (1-x_1-x_2) \int \frac{d^4 k_1}{(2\pi)^4} \frac{[g^{\alpha\mu }-k_1^\alpha k_1^\mu /m_W^2]}{(k_1^2-m_W^2)(k_1^2-m_b^2)[(k_1+k_2)^2-m_\rho^2]} \frac{2 g_{\mu\nu}}{(m_x^{\prime 2})^2}\, ,\,\,
\label{Intk1_xi}
\end{eqnarray}
where the integration measure over the Feynman parameters $x_j$ and $s_i$ are suppressed.
The combination of the denominator factors introduces the Feynman parameters $z_i$ ($i=1,...3$), which leads to 
\begin{eqnarray}\label{II1_xi}
{\rm II}^\prime_{1} = \frac{-i}{16\pi^2} \frac{x_2 (1-x_1-x_2)}{x_1^2 (1-x_1)^2} \Gamma(5)\int \prod^3_{j=1} d z_j \frac{2z_1[g^\alpha_\nu-k_1^\alpha k_{1\nu} /m_W^2]}{D^{\prime 5}} \, ,
\end{eqnarray}
where
\begin{eqnarray}\label{Dp_exp}
D^\prime &=& z_1\Big[k_1^2+\frac{2x_1 x_2}{x_1(1-x_1)}k_1\cdot k_2 + \frac{x_2(1-x_2)}{x_1(1-x_1)}k_2^2 - \frac{m_\sigma^2}{x_1(1-x_1)}\Big] +z_2 \Big[(k_1+k_2)^2 -m_\rho^2 \Big]\nonumber\\
&& +z_3(k_1^2-m_W^2)+(1-z_1-z_2-z_3)(k_1^2-m_b^2)
\end{eqnarray}

After the internal momentum translation: $k_1\rightarrow k_1- c_2 k_2$ where $c_2 = [x_2 z_1+(1-x_1)z_2]/(1-x_1)$, we can integrate out $k_1$, resulting in:
\begin{eqnarray}\label{II1p_xi}
{\rm II}^\prime_{1} = \frac{1}{(16\pi^2)^2} \int \prod_{j=1}^3 d z_j \frac{ x_2 (1-x_1-x_2)z_1}{x_1^2 (1-x_1)^2} \Big\{ \frac{4 (g^\alpha_\nu - c_2^2 k_{2}^\alpha k_{2\nu}/m_W^2)}{B_2^3 [k_2^2 - \Delta^\prime]^3} - \frac{1}{m_W^2} \frac{g^\alpha_\nu}{B_2^2 [k_2^2 - \Delta^\prime]^2} \Big\}\, ,\nonumber\\
\end{eqnarray}
where $B_2$ is defined in Eq.~(\ref{B2Func}) and
\begin{eqnarray}
\Delta^\prime &=& \frac{z_2}{B_2}m_\rho^2 + \frac{z_3}{B_2}m_W^2 + \frac{1-z_1-z_2-z_3}{B_2} m_b^2 + \frac{z_1 m_\sigma}{B_2x_1(1-x_1)}\, .
\end{eqnarray}
We now turn to the integration over $k_2$,
\begin{eqnarray}
{\rm III}^\prime_{1} &=& \frac{1}{(16\pi^2)^2} \frac{x_2 (1-x_1-x_2) z_1}{x_1^2 (1-x_1)^2} \int \frac{d^4 k_2}{(2\pi)^4} \frac{(g^{\beta\nu}-k_2^\beta k_2^\nu / m_W^2)}{(k_2^2-m_W^2)(k_2^2-m_a^2)} \nonumber\\
&& \Big\{ \frac{4 (g^\alpha_\nu - c_2^2 k_{2}^\alpha k_{2\nu}/m_W^2)}{B_2^3 [k_2^2 - \Delta^\prime]^4} - \frac{1}{m_W^2} \frac{g^\alpha_\nu}{B_2^2 [k_2^2 - \Delta^\prime]^3} \Big\}\, ,
\end{eqnarray}
where the integration measures for the Feynman parameters are also suppressed. The integration over $k_2$ can be performed with the help of the Feynman parameters $y_i$,
\begin{eqnarray}\label{I31Final_xi}
{\rm III}^\prime_{1} &=& \frac{1}{(16\pi^2)^2} \frac{ x_2 (1-x_1-x_2)z_1}{x_1^2(1-x_1)^2} \int d y_1 d y_2 \int \frac{d^4 k_2}{(2\pi)^4} (g^{\beta\nu}- k_2^\beta k_2^\nu/m_W^2)\nonumber\\
&& \Big\{ \frac{\Gamma(5)}{\Gamma(3)}\frac{4y_1^2}{B_2^3} \frac{(g_\nu^\alpha - c_2^2 k_2^\alpha k_{2\nu}/m_W^2)}{(k_2^2 - [m_\rho^2 \Sigma^\prime]/[B_2 x_1 (1-x_1)])^5}\nonumber\\
&& - \Gamma(4)\frac{y_1}{m_W^2 B_2^2} \frac{g^\alpha_\nu}{(k_2^2 - [m_\rho^2\Sigma^\prime]/[B_2 x_1 (1-x_1)])^4} \Big\}\, \nonumber\\
&=& \frac{-i}{(16\pi^2)^3}{[x_2 (1-x_1-x_2)z_1]} \frac{g^{\alpha\beta}}{4} \Big\{ y_1^2[\frac{16 x_1 (1-x_1)}{m_\rho^6\Sigma^{\prime 3}} + \frac{4(c_2^2+1)}{m_W^2 m_\rho^4 B_2 \Sigma^{\prime 2}}\nonumber\\
&&  + \frac{12 c_2^2}{m_W^4 m_\rho^2 B_2^2 x_1(1-x_1) \Sigma^\prime }] +y_1 [\frac{4}{m_W^2 m_\rho^4 \Sigma^{\prime 2}} + \frac{2}{m_W^4 m_\rho^2 B_2 x_1 (1-x_1) \Sigma^\prime}] \Big\}\, .
\end{eqnarray}
where
\begin{eqnarray}
\Sigma^\prime &=& x_1(1-x_1)[y_1 z_2 + (y_1 z_3 + y_2 B_2)\frac{m_W^2}{m_\rho^2} + y_1 (1-z_1-z_2-z_3)\frac{m_b^2}{m_\rho^2} +(1-y_1-y_2) B_2 \frac{m_a^2}{m_\rho^2}]\nonumber\\
&& + y_1 z_1 \frac{m_\sigma^2}{m_\rho^2} \,.
\end{eqnarray}

\subsubsection{Integration of ${\rm I}^\prime_{12}$ Over $k_1$ and $k_2$}
The integration over $k_1$ for ${{\rm I}^\prime_{12}}$ is defined as
\begin{eqnarray}
{\rm II}^\prime_2 = \frac{-i}{16\pi^2} x_2 (1-x_1-x_2) \int\frac{d^4 k_1}{(2\pi)^4}
\frac{(g^{\alpha\mu}-k_1^\alpha k_1^\mu / m_W^2)}{(k_1^2-m_W^2)(k_1^2-m_b^2)[(k_1+k_2)^2-m_\rho^2]} \frac{2 N_{\mu\nu}}{(m^{\prime 2}_x)^3} \, .\,\,
\end{eqnarray}
With the same Feynman parameters $z_i$ as that in Eq.~(\ref{II1}) and the same internal momentum shift $k_1\to k_1 -c_2 k_2$, ${\rm II}^\prime_2$ can be transformed into
\begin{eqnarray}
{\rm II}^\prime_2 = \frac{i}{16\pi^2} \frac{ x_2 (1-x_1-x_2)}{x_1^3 (1-x_1)^3}{\Gamma(6)} z_1^3 \int\frac{d^4 k_1}{(2\pi)^4} \frac{N^{\prime \alpha}_\nu}{D^{\prime 6}}\, ,
\end{eqnarray}
where $D^\prime$ is defined in Eq.~(\ref{Dp_exp}) and $N'^\alpha_\nu$ is the same as that in Eqs.~(\ref{Def_N}) and (\ref{Def_dd}).
The integral over $k_1$ can be worked out with the result given by
\begin{eqnarray}
{\rm II}^\prime_2&=&\frac{1}{(16\pi^2)^2}\frac{x_2(1-x_1x_2)z_1^2}{x_1^3(1-x_1)^3}\Big\{
-\frac{6 d_1d_2 k_2^\alpha k_{2\nu}}{B_2^4[k_2^2-\Delta^\prime]^4}
+\frac{6 d_1d_2c_2^2 k_2^2 k_2^\alpha k_{2\nu}}{B_2^4 m_W^2[k_2^2-\Delta^\prime]^4}
-\frac{(1-2\,x_1)(2\,x_1)g_{\nu}^\alpha}{ B_2^3[k_2^2-\Delta^\prime]^3}\nonumber\\
&&+\frac{(1-2\,x_1)(2\,x_1)c_2^2k_{2\nu}k_{2}^\alpha}{B_2^3 m_W^2[k_2^2-\Delta^\prime]^3}
+\frac{(1-2\,x_1)d_2c_2k_{2\nu}k_{2}^\alpha}{B_2^3 m_W^2 [k_2^2-\Delta^\prime]^3}
+\frac{4 (1-2\,x_1)d_2c_2 k_{2\nu}k_{2}^\alpha}{B_2^3 m_W^2[k_2^2-\Delta^\prime]^3}\nonumber\\
&&+\frac{(2\,x_1)d_1c_2g_{\nu}^\alpha k_{2}^2}{B_2^3 m_W^2 [k_2^2-\Delta^\prime]^3}
+\frac{(2\,x_1)d_1c_2 k_{2\nu}k_2^\alpha}{B_2^3 m_W^2 [k_2^2-\Delta^\prime]^3}
+\frac{d_1d_2k_{2\nu}k_2^\alpha}{B_2^3 m_W^2[k_2^2-\Delta^\prime]^3}
+\frac{3(1-2\,x_1)(2\,x_1)g_{\nu}^\alpha}{2 B_2^2 m_W^2 [k_2^2-\Delta^\prime]^2}
\Big\}\;, \nonumber\\
\end{eqnarray}
On the basis of ${\rm II}_2^\prime$, we can write down the expression for the integration over $k_2$
by appending the rest propagators, and perform the Feynman parametrization with $y_i$ as that in Eq.~(\ref{I31Final_xi}) to transform the expression into
\begin{eqnarray}
{\rm III}_2^\prime &=& \int\frac{d^4k_2}{(2\pi)^4}
\frac{({g^{\beta\nu}-k_2^\beta k_2^\nu/m_W^2})}{(k_2^2-m_W^2)(k_2^2-m_a^2)}{\rm II}^\prime_2 \nonumber\\
&=&\frac{1}{(16\pi^2)^2}\frac{x_2(1-x_1-x_2)z_1^2}{ x_1^3(1-x_1)^3}\int\frac{d^4k_2}{(2\pi)^4}\Big\{
-\frac{\Gamma(6)d_1 d_2 y_1^3}{B_2^4}\frac{{k_2^\alpha k_2^\beta(1-k_2^2/m_W^2)}}{ [k_2^2-(m_\rho^2\Sigma^\prime)/ (x_1(1-x_1)B_2)]^6}\nonumber\\
&&+\frac{\Gamma(6) d_1 d_2 c_2^2 y_1^3}{B_2^4 m_W^2}\frac{k_2^\alpha k_2^\beta k_2^2(1-k_2^2/ m_W^2)}{[k_2^2-(m_\rho^2\Sigma^\prime)/ (x_1(1-x_1)B_2)]^6}
\nonumber\\
&&
-\frac{\Gamma(5)(1-2x_1)(2x_1)y_1^2}{\Gamma(3) B_2^3}\frac{(g^{\alpha\beta}-{k_{2}^\alpha k_2^\beta/m_W^2}) }{ [k_2^2-(m_\rho^2 \Sigma^\prime)/ (x_1(1-x_1)B_2)]^5}\nonumber\\
&&+\frac{\Gamma(5)(1-2x_1)(2x_1)c_2^2y_1^2}{\Gamma(3) B_2^3 m_W^2}\frac{k_2^\alpha k_2^\beta (1-{k_2^2/ m_W^2})}{ [k_2^2-(m_\rho^2\Sigma^\prime)/ (x_1(1-x_1)B_2)]^5}\nonumber\\
&&+\frac{\Gamma(5)(1-2x_1)d_2c_2y_1^2}{ \Gamma(3) B_2^3 m_W^2}\frac{k_2^\alpha k_2^\beta (1-{k_2^2/ m_W^2})}{[k_2^2-(m_\rho^2\Sigma^\prime)/ (x_1(1-x_1)B_2)]^5}\nonumber\\
&&+\frac{4\Gamma(5)(1-2x_1) d_2 c_2 y_1^2}{\Gamma(3) B_2^3 m_W^2}\frac{k_2^\alpha k_2^\beta (1-{k_2^2/ m_W^2})}{[k_2^2-(m_\rho^2\Sigma^\prime)/ (x_1(1-x_1)B_2)]^5}\nonumber\\
&&+\frac{\Gamma(5)(2x_1)d_1c_2y_1^2}{\Gamma(3) B_2^3 m_W^2}\frac{k_2^2(g^{\alpha\beta}-{k_2^\alpha k_2^\beta / m_W^2})}{[k_2^2-(m_\rho^2\Sigma^\prime)/ (x_1(1-x_1)B_2)]^5}\nonumber\\
&&+\frac{\Gamma(5)(2x_1) d_1 c_2 y_1^2}{\Gamma(3)B_2^3 m_W^2}\frac{ k_2^\alpha k_2^\beta(g^{\alpha\beta}-{k_2^2/m_W^2})}{[k_2^2-(m_\rho^2\Sigma^\prime)/ (x_1(1-x_1)B_2)]^5}
\nonumber\\
&&
+\frac{\Gamma(5) d_1 d_2 y_1^2}{\Gamma(3) B_2^3 m_W^2}\frac{ k_2^\alpha k_2^\beta(1-{k_2^2/m_W^2})
}{[k_2^2-(m_\rho^2\Sigma^\prime)/ (x_1(1-x_1)B_2)]^5}\nonumber\\
&&+\frac{9(1-2x_1)(2x_1)y_1 }{B_2^2 m_W^2}\frac{g^{\alpha\beta}-{k_2^\alpha k_2^\beta / m_W^2}} {[k_2^2-(m_\rho^2\Sigma^\prime)/ (x_1(1-x_1)B_2)]^4}
\Big\}\;.\nonumber\\
\end{eqnarray}

Finally, the integration over $k_2$ results in
\begin{eqnarray}
{\rm III_2}^\prime &=&\frac{i}{(16\pi^2)^3}[x_2 (1-x_1-x_2) z_1^2] \frac{g^{\alpha\beta}}{4}\Big\{
d_1 d_2 y_1^3(\frac{4}{B_2\Sigma^{\prime 3} m_\rho^6}+ \frac{6 }{x_1(1-x_1)B_2^2\Sigma^{\prime 2} m_\rho^4 m_W^2})\nonumber\\
&& + d_1 d_2 c_2^2 y_1^3 (\frac{6}{x_1(1-x_1)B_2^2\Sigma^{\prime 2} m_\rho^4 m_W^2}+\frac{24}{x_1^2 (1-x_1)^2 B_2^3 \Sigma^\prime m_\rho^2 m_W^4}) \nonumber\\
&& + (2x_1)(1-2x_1)y_1^2(\frac{4}{\Sigma^{\prime 3} m_\rho^6}+\frac{1}{x_1(1-x_1) B_2 \Sigma^{\prime 2} m_\rho^4 m_W^2})\nonumber\\
&& + (2x_1) (1-2x_1) c_2^2 y_1^2 (\frac{1}{x_1 (1-x_1) B_2 \Sigma^{\prime 2} m_\rho^4 m_W^2}+ \frac{3}{x_1^2 (1-x_1)^2 B_2^2 \Sigma^\prime m_\rho^2 m_W^4})\nonumber\\
&& + (1-2x_1) d_2 c_2 y_1^2 (\frac{1}{x_1 (1-x_1) B_2 \Sigma^{\prime 2} m_\rho^4 m_W^2}+ \frac{3}{x_1^2 (1-x_1)^2 B_2^2 \Sigma^\prime m_\rho^2 m_W^4})\nonumber\\
&& + (1-2x_1) d_2 c_2 y_1^2 (\frac{4}{x_1 (1-x_1) B_2 \Sigma^{\prime 2} m_\rho^4 m_W^2}+\frac{12}{x_1^2 (1-x_1)^2 B_2^2 \Sigma^\prime m_\rho^2 m_W^4})\nonumber\\
&& + (2x_1) d_1 c_2 y_1^2 (\frac{4}{x_1 (1-x_1) B_2 \Sigma^{\prime 2} m_\rho^4 m_W^2}+ \frac{3}{x_1^2 (1-x_1)^2 B_2^2 \Sigma^\prime m_\rho^2 m_W^4})\nonumber\\
&& + (2x_1) d_1 c_2 y_1^2 (\frac{1}{x_1 (1-x_1) B_2 \Sigma^{\prime 2} m_\rho^4 m_W^2}+ \frac{3}{x_1^2(1-x_1)^2 B_2^2 \Sigma^\prime m_\rho^2 m_W^4})\nonumber\\
&& + d_1 d_2 y_1^2 (\frac{1}{x_1 (1-x_1) B_2 \Sigma^{\prime 2} m_\rho^4 m_W^2}+\frac{3}{x_1^2 (1-x_1)^2 B_2^2 \Sigma^\prime m_\rho^2 m_W^4})\nonumber\\
&& + (1-2x_1) (2x_1) y_1 (\frac{6}{x_1 (1-x_1) \Sigma^{\prime 2} m_\rho^4 m_W^2}+\frac{3}{ x_1^2 (1-x_1)^2 B_2 \Sigma^\prime m_\rho^2 m_W^4})
\Big\}\;.
\end{eqnarray}

\subsection{Final Results}
The final analytic formula for ${\cal I}_2^\prime$ is obtained by summing up that for ${\rm III}_1^\prime$ and ${\rm III}_2^\prime$, given by
\begin{eqnarray}\label{result_xi}
{\cal I}^\prime_2 = \frac{1}{2}(16\pi^2)^3 m_\rho^4 m_W^4 \gamma_\beta \gamma_\alpha ({\rm III}^\prime_1 + {\rm III}^\prime_2)
= \frac{i}{2} (\frac{m_W^4}{m_\rho^4}{\cal I}^\prime_{22} + \frac{m_W^2}{m_\rho^2}{\cal I}^\prime_{21} + {\cal I}^\prime_{20} )\,,
\end{eqnarray}
where
\begin{eqnarray}
{\cal I}^\prime_{20} &=& \frac{x_2 (1-x_1-x_2) z_1}{\Sigma^{\prime 3}} \Big\{ - 16 x_1 (1-x_1) y_1^2 + \frac{4 d_1 d_2\, y_1^3 z_1}{B_2} + 4(1-2 x_1)(2x_1)y_1^2 z_1 \Big\}\, ,
\end{eqnarray}
\begin{eqnarray}
{\cal I}^\prime_{21} &=& -\frac{x_2 (1-x_1-x_2) z_1}{\Sigma^{\prime 2}} \Big\{ \frac{4 y_1^2 (c_2^2+1)}{B_2} + 4 y_1 \Big\} + \frac{x_2 (1-x_1-x_2) z_1^2}{x_1 (1-x_1 )\Sigma^{\prime 2}}\Big\{ \frac{6 d_1 d_2\, y_1^3}{B_2^2}\nonumber\\
&&  + \frac{6 d_1 d_2 c_2^2 y_1^3}{B_2^2} + \frac{(1-2x_1)(2x_1)y_1^2}{B_2} + \frac{(1-2 x_1)(2 x_1) c_2^2 y_1^2}{B_2} + \frac{(1-2 x_1)d_2 c_2 y_1^2}{B_2}  \nonumber\\
&& + \frac{4 (1-2 x_1) d_2 c_2 y_1^2}{B_2} + \frac{4 (2 x_1) d_1 c_2 y_1^2 }{B_2} + \frac{(2 x_1) d_1 c_2 y_1^2}{B_2} +\frac{ d_1 d_2 y_1^2}{B_2} + 6 (1-2 x_1)(2x_1)y_1 \Big\} \, ,\nonumber\\
\end{eqnarray}
\begin{eqnarray}
{\cal I}^\prime_{22} &=& -\frac{x_2 (1-x_1- x_2)z_1 }{x_1 (1-x_1) \Sigma^\prime} \Big\{ \frac{12 y_1^2 c_2^2}{B_2^2} + \frac{2 y_1}{B_2} \Big\} + \frac{x_2 (1-x_1-x_2) z_1^2}{x_1^2 (1-x_1)^2 \Sigma^\prime} \Big\{ \frac{24 d_1 d_2 c_2^2 y_1^3}{B_2^3} \nonumber\\
&&  + \frac{3(1-2 x_1)(2 x_1) c_2^2 y_1^2}{B_2^2} + \frac{3 (1-2 x_1) d_2 c_2 y_1^2}{B_2^2} + \frac{12 (1-2 x_1) d_2 c_2 y_1^2}{B_2^2} +\frac{3 (2 x_1) d_1 c_2 y_1^2}{B_2^2}\nonumber\\
&& + \frac{3 (2 x_1) d_1 c_2 y_1^2}{B_2^2} + \frac{3 d_1 d_2 y_1^2}{B_2^2} +\frac{3 (1-2x_1)(2x_1) y_1}{B_2} \Big\}\, .
\end{eqnarray}
Note that the 9-dimensional integration measure for the Feynman parameters $x_i,\, y_i,\, z_i$ previously suppressed is defined by
\begin{eqnarray}\label{measure_xi}
{\rm measure} = \int^1_0 ds_2 \int^1_0 ds_3 \int^1_0 d x_1 \int^{1-x_1}_0 dx_2 \int^1_0 dy_1 \int^{1-y_1}_0 dy_2 \int^1_0 dz_1 \int^{1-z_1}_0 dz_2 \int^{1-z_1-z_2}_0 dz_3\,. \nonumber\\
\end{eqnarray}
As mentioned before, ${\cal I}^{\prime\prime}_2$ can be simply obtained by exchanging the charged scalar masses $m_{H_1} \leftrightarrow m_{H_2}$ in Eq.~(\ref{result_xi}). This completes our analytical derivation of the integral ${\cal I}_2$.

For the remaining 9-dimensional Feynman parameter integrations in ${\cal I}_2$, we also use the three packages as in the $\kappa_2$ part calculation: {\sf Mathematica}(Global Adaptive), {\sf SecDec-2.1.4}~\cite{SecDec} and VEGAS in {\sf GSL}~\cite{GSL}, in order to make a cross check. Consequently, all of them give the essentially the same result within errors. For the particle spectrum of the benchmark point listed in Eq.~(\ref{Eq_BMpoint}), the three-loop integration ${\cal I}_2$ is given by,
\begin{eqnarray}
{\cal I}_2 = 4.15 i.
\end{eqnarray}
Together with ${\cal I}_1 = 2.16 i$ as calculated in the previous section, we can predict the neutrino mass matrix numerically by taking various possible values of $\xi$, and our results are given by
\begin{eqnarray}
\xi=0.5:  &\quad\quad &  \quad m_{\nu} =
\left(
\begin{matrix}
{\mathcal O}(10^{-3}) & {\mathcal O}(10^{-3}) &   1.52 \\
{\mathcal O}(10^{-3}) &  3.14 &  4.74 \\
1.52 & 4.74  & 4.42
 \end{matrix}
\right)\times 10^{-13}~~\mathrm{GeV}.
\label{xi05}
\end{eqnarray}

\begin{eqnarray}
\xi=0.8:  &\quad\quad &  \quad m_{\nu} =
\left(
\begin{matrix}
{\mathcal O}(10^{-3}) & {\mathcal O}(10^{-3}) &   1.81 \\
{\mathcal O}(10^{-3}) &  3.74 &  5.64 \\
1.81 & 5.64  & 5.25
 \end{matrix}
\right)\times 10^{-13}~~\mathrm{GeV}.
\label{xi08}
\end{eqnarray}

\begin{eqnarray}
\xi=1:  &\quad\quad &  \quad m_{\nu} =
\left(
\begin{matrix}
{\mathcal O}(10^{-3}) & {\mathcal O}(10^{-3}) &   2.00 \\
{\mathcal O}(10^{-3}) &  4.13 &  6.23 \\
2.00 & 6.23  & 5.80
 \end{matrix}
\right)\times 10^{-13}~~\mathrm{GeV}.
\label{xi1}
\end{eqnarray}

\begin{eqnarray}\label{Extreme}
\xi=5:  &\quad\quad &  \quad m_{\nu} =
\left(
\begin{matrix}
{\mathcal O}(10^{-3}) & {\mathcal O}(10^{-3}) &   5.80 \\
{\mathcal O}(10^{-3}) &  12.0 &  18.1 \\
5.80 & 18.1  & 16.9
 \end{matrix}
\right)\times 10^{-13}~~\mathrm{GeV}.
\label{xi5}
\end{eqnarray}
Note that Eq.~(\ref{Extreme}) can be regarded as the extreme case allowed by the naturalness argument~\cite{Nebot:2007bc}. 
In sum, we see that the predicted neutrino mass matrix elements are typically smaller than the realistic values up to two orders 
of magnitude for the benchmark point shown in the first version of Ref.~\cite{Gustafsson:2012vj} before the publication of its Erratum.

\end{document}